\documentclass[twocolumn]{aastex62}

\usepackage{amsmath,amsfonts,amssymb,comment,gensymb,courier,color,xcolor,float}
\usepackage{natbib}
\usepackage{hyperref}
\usepackage{graphicx}

\shorttitle{The Dragonfly Nearby Galaxies Survey. V.}
\shortauthors{Cohen et al.}
	
\begin{document}

\title{The Dragonfly Nearby Galaxies Survey. V. HST/ACS Observations of 23 Low Surface Brightness Objects in the Fields of NGC\,1052, NGC\,1084, M\,96, and NGC\,4258}


\author[0000-0001-5487-2494]{Yotam Cohen}
\affiliation{Department of Astronomy, Yale University, New Haven, CT 06511, USA}

\author{Pieter van Dokkum}
\affiliation{Department of Astronomy, Yale University, New Haven, CT 06511, USA}

\author{Shany Danieli}
\affil{Department of Astronomy, Yale University, New Haven, CT 06511, USA}
\affil{Department of Physics, Yale University, New Haven, CT 06520, USA}
\affil{Yale Center for Astronomy and Astrophysics, Yale University, New Haven, CT 06511, USA}

\author{Aaron J. Romanowsky}
\affil{Department of Physics \& Astronomy, San Jos\'{e} State University, One Washington Square, San Jose, CA 95192, USA}
\affil{University of California Observatories, 1156 High Street, Santa Cruz, CA 95064, USA}

\author{Roberto Abraham}
\affil{Department of Astronomy \& Astrophysics, University of Toronto, 50 St. George Street, Toronto, Ontario M5S 3H4, Canada}
\affil{Dunlap Institute for Astrophysics, University of Toronto, 50 St. George Street, Toronto, Ontario M5S 3H4, Canada}

\author{Allison Merritt}
\affiliation{Max-Planck-Institut f\"{u}r Astronomie, K\"{o}nigstuhl 17, D-69117 Heidelberg, Germany}

\author{Jielai Zhang}
\affil{Department of Astronomy \& Astrophysics, University of Toronto, 50 St. George Street, Toronto, Ontario M5S 3H4, Canada}
\affil{Dunlap Institute for Astrophysics, University of Toronto, 50 St. George Street, Toronto, Ontario M5S 3H4, Canada}

\author{Lamiya Mowla}
\affil{Department of Astronomy, Yale University, New Haven, CT 06511, USA}

\author{J.~M. Diederik Kruijssen}
\affiliation{Astronomisches Rechen-Institut, Zentrum f\"{u}r Astronomie der Universit\"{a}t Heidelberg, M\"{o}nchhofstra\ss e 12-14, 69120 Heidelberg, Germany}

\author{Charlie Conroy}
\affiliation{Harvard-Smithsonian Center for Astrophysics, 60 Garden Street, Cambridge, MA}

\author[0000-0003-4235-3595]{Asher Wasserman}
\affil{Department of Astronomy \& Astrophysics, University of California-Santa Cruz, Santa Cruz, CA 95064, USA}

\begin{abstract}
We present \textit{HST}/ACS imaging of twenty-three very low surface brightness ($ \mu_{e,V} \sim 25 - 27.5 $) galaxies detected in the fields of four nearby galaxy groups.
These objects were selected from deep optical imaging obtained with the Dragonfly Telephoto Array.
Seven are new discoveries, while most of the others had been identified previously in visual surveys of deep photographic plates and more recent surveys.
Few have previously been studied in detail.
From the ACS images, we measure distances to the galaxies using both the tip of the red giant branch method and the surface brightness fluctuations method.
We demonstrate that the two methods are consistent with each other in the regime where both can be applied.
The distances to 15 out of 20 galaxies with stable measurements are consistent with that of the targeted group within errors.
This suggests that assuming group membership based solely on projected proximity is $\sim 75$\,\% successful in this regime.
The galaxies are nearly round, with a median axis ratio of 0.85, and visually resemble dwarf spheroidal galaxies.
The objects have a range of sizes, from $R_e = 0.4$\,kpc to $R_e = 1.8$\,kpc, with a median $\langle R_e \rangle = 1.0$\,kpc.
They range in luminosity from $M_V = -11.4$ to $M_V = -15.6$, with a median $\langle M_V \rangle = -12.4$.
Galaxies with $R_e\sim 1$\,kpc and $M_V\sim -12$  are fairly rare in the Local Group but we find many of them in this relatively small sample.
Four of the objects fall in the class of ultra diffuse galaxies (UDGs), with $R_e>1.5$\,kpc and $\mu_{0,V}>24 \ \mathrm{mag \ arcsec^{-2}}$, including the recently identified dark matter deficient galaxy NGC\,1052-DF2.
\end{abstract}


\section{Introduction} \label{sec:intro}

While studies of low surface brightness (LSB) galaxies date back several decades \citep{bothun87,impey96}, advances in instrumentation are now allowing us to push down to ever dimmer limits and reveal the diversity of the LSB universe.

We have been taking deep, wide field images of the sky using a robotic, refracting, low-surface brightness optimized telescope called the Dragonfly Telephoto Array \citep{dragonfly}.
Using this telescope, a substantial population of ultra--diffuse galaxies (UDGs) featuring extremely low surface brightnesses ($\mu_{0,g} > 24\, \mathrm{mag \, arcsec^{-2}}$) and large effective radii ($R_{\rm eff} > 1.5\, \mathrm{kpc}$) was identified by \cite{vd2015a} in the Coma cluster.
These UDGs have drawn significant attention in recent years and subsequent searches have found similar objects in other galaxy clusters including Virgo \citep{mihos15} and Fornax \citep{munoz15}. The formation of UDGs is a topic of vigorous debate in the literature \citep[see, e.g.,][]{mowla17,BT16,dicintio17}.

Dragonfly is also contributing to our knowledge of galaxies in groups and the general field, as in most $\sim 6$ square degree images of bright nearby galaxies there are low surface brightness objects that may be dwarf satellites of the primary galaxy \citep[see, e.g.,][]{kk1998,kks2000}. 
This low surface brightness approach complements searches that use star counts \citep[e.g.][]{Belokurov:2007,Martin:2013,Smercina:2018} as it can be used effectively at relatively large distances \citep[see][]{danieli18}.

We have obtained images of a sample of nearby galaxies in the context of the Dragonfly Nearby Galaxy Survey \citep[DNGS;][]{Merritt:2016,merritt2016}.
The galaxies in this survey are the nearest objects in bins of absolute magnitude, with additional constraints on Galactic cirrus emission and visibility from the New Mexico Skies Observatory.
The first galaxy that we observed was the nearby massive spiral M\,101 \citep{vDm101}.
We identified seven previously-unknown low LSB objects \citep{merritt2014}.
Follow-up \textit{HST}/ACS imaging was used to measure distances using the tip of the red giant branch (TRGB) method, and it turned out that, surprisingly, only three out of the seven objects are in fact dwarf galaxies at the distance of M\,101 \citep{danieli2016}.
The other four are most likely UDGs associated with the background NGC\,5485 group at a distance of $\sim 27$\,Mpc \citep{merritt2016}.

In this paper we describe \textit{HST}/ACS observations of 23 other low surface brightness objects in four of the DNGS fields. 
More than half of the galaxies had already been identified in previous searches \citep[][]{kks2000,TT2002,kkh2011,kmk2013,kkm2014,krz2015,muller18}, but they had not been studied in detail and none had accurately measured distances.
In this work we present these new ACS data, and measure structural parameters and distances of the galaxies.
One object in this sample is NGC\,1052-DF2, which has already been the subject of several papers \citep{df2nature,df2apj,df2rn,df2stream,martin18,laporte18,nusser18,wasserman18}.

\section{Observations and Data Reduction} \label{sec:observations}

\subsection{Identification with Dragonfly}

The diffuse galaxies examined in this work were identified as candidates in deep images taken with Dragonfly in a 10-lens configuration.\footnote{At the time of writing, Dragonfly has 48 lenses.}
The images are centered on bright galaxies in four nearby groups: NGC\,1052, NGC\,3384 (in the M\,96 group), NGC\,4258, and NGC\,1084.
The deep frames are coadds of many 600 second exposures.
Typical total exposure times were between 15 and 25 hours per field \citep[see][]{Merritt:2016} in $g$ and $r$ filters.

\begin{figure*}[]
\label{fields}
\gridline{\fig{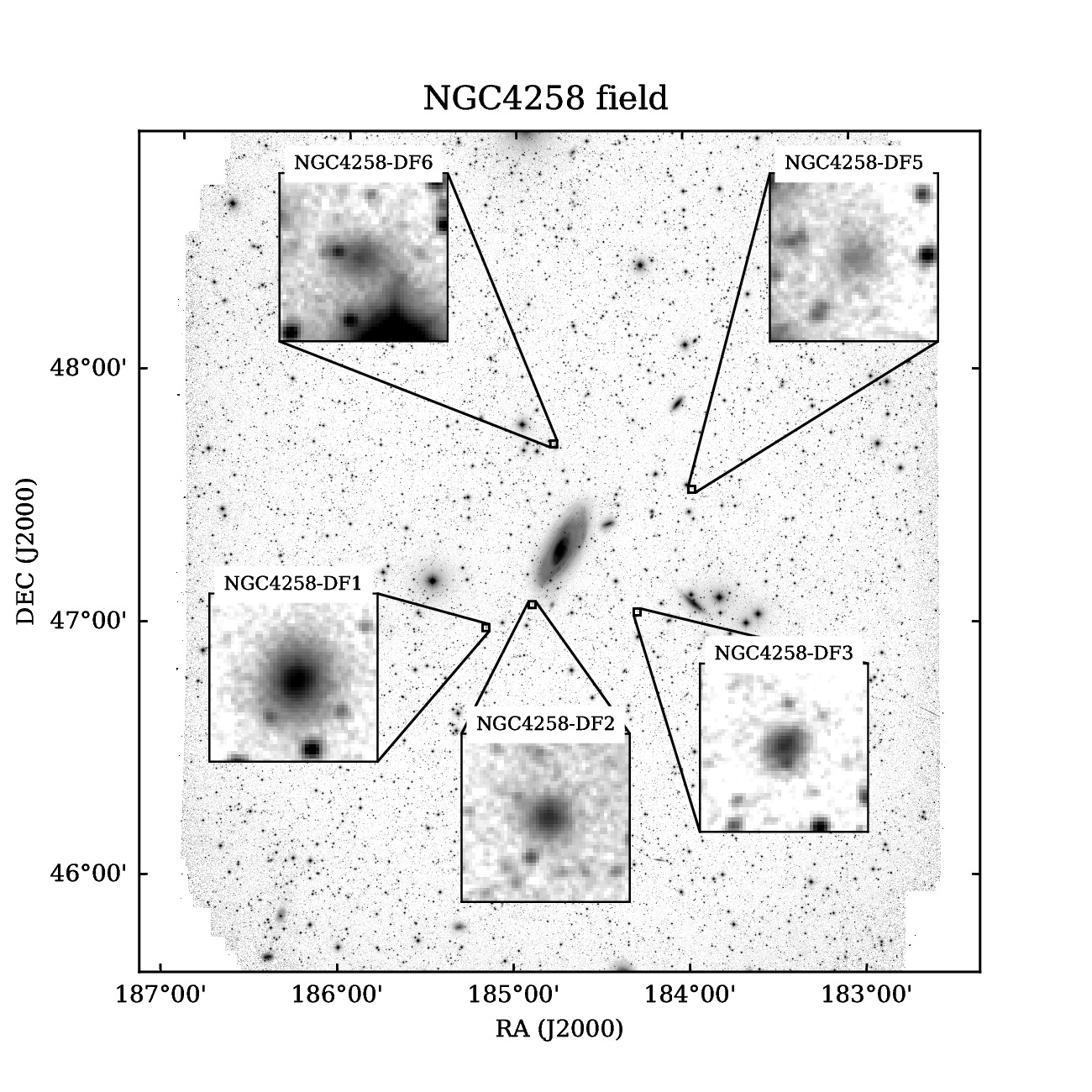}{0.5\textwidth}{}
		\fig{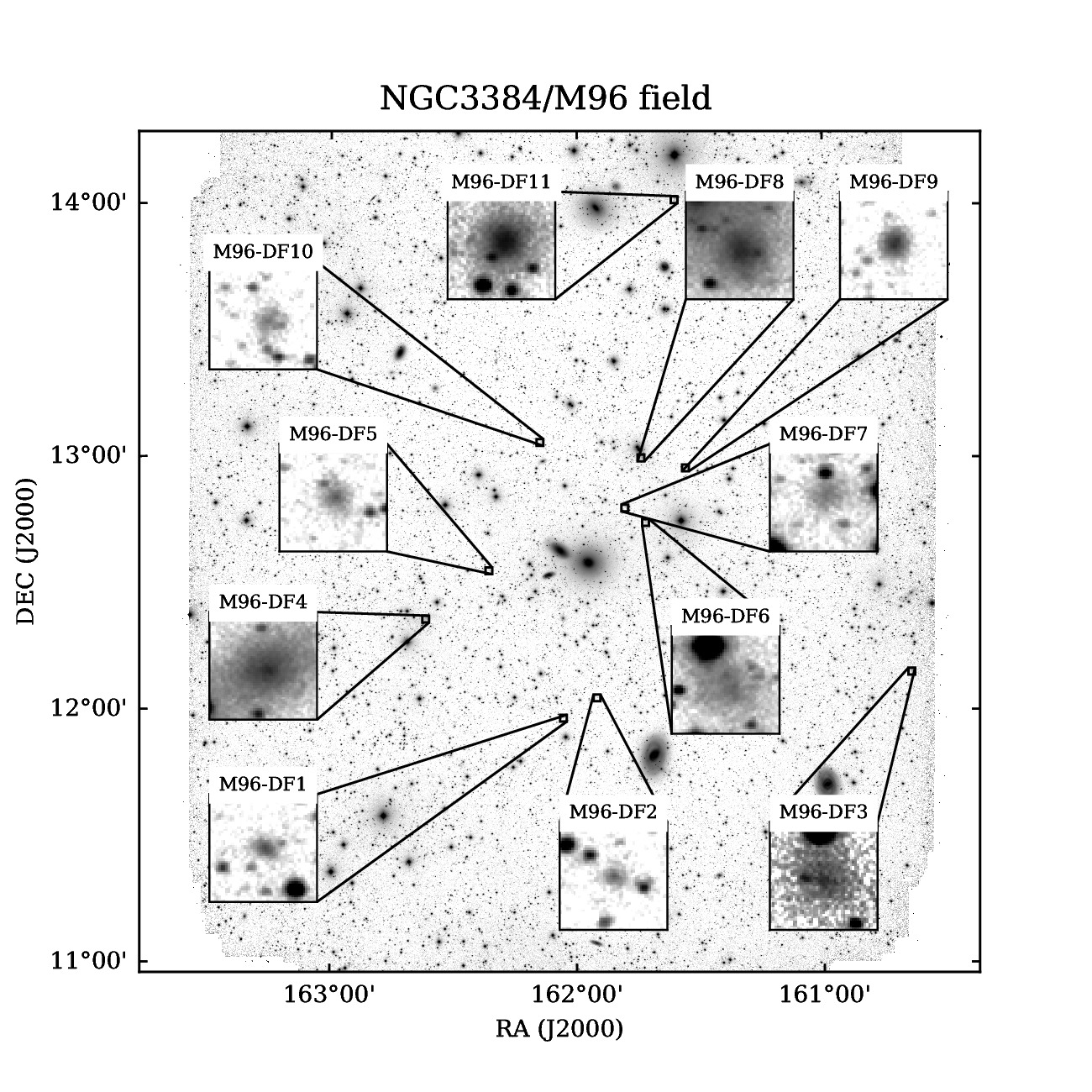}{0.5\textwidth}{}}
\gridline{\fig{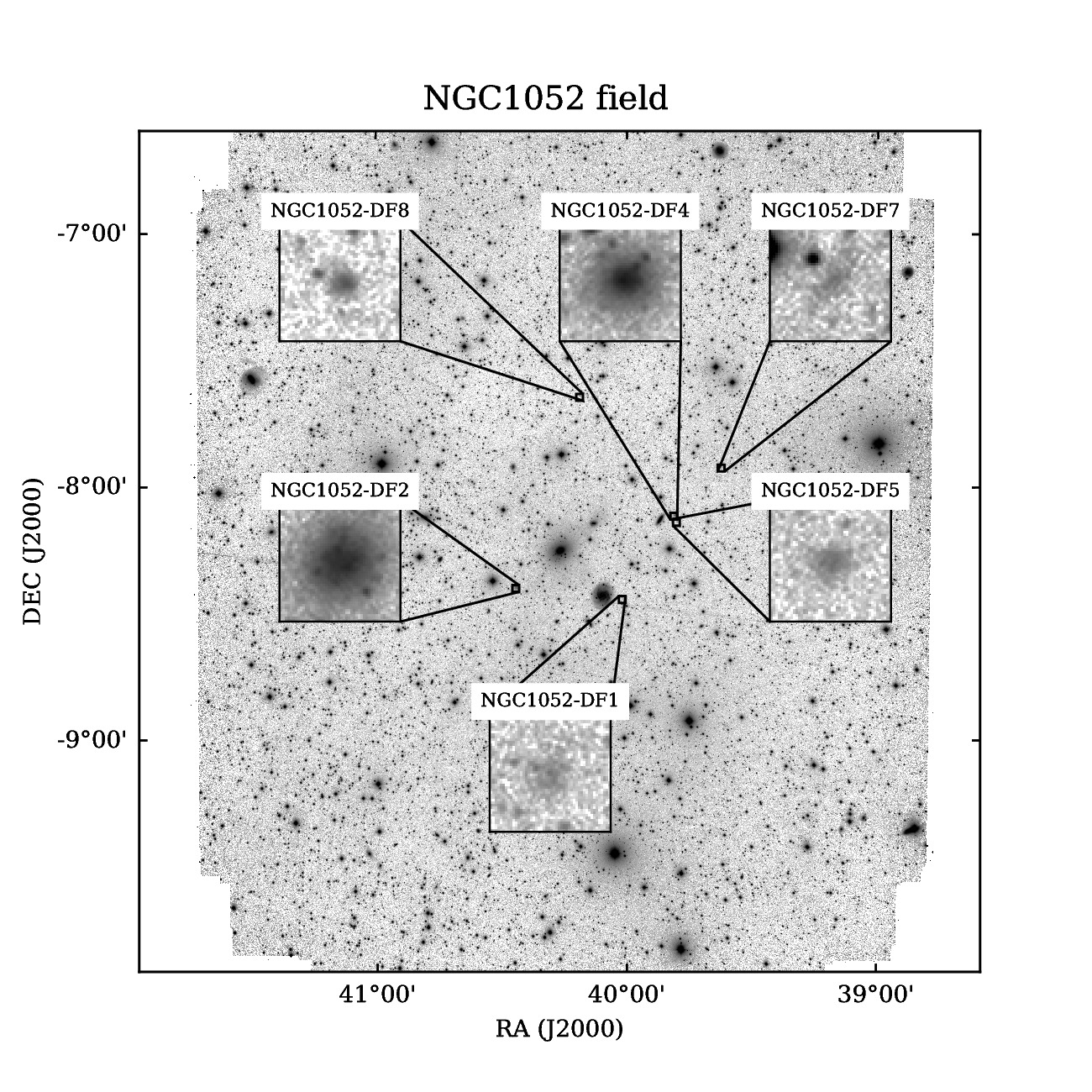}{0.5\textwidth}{}
		\fig{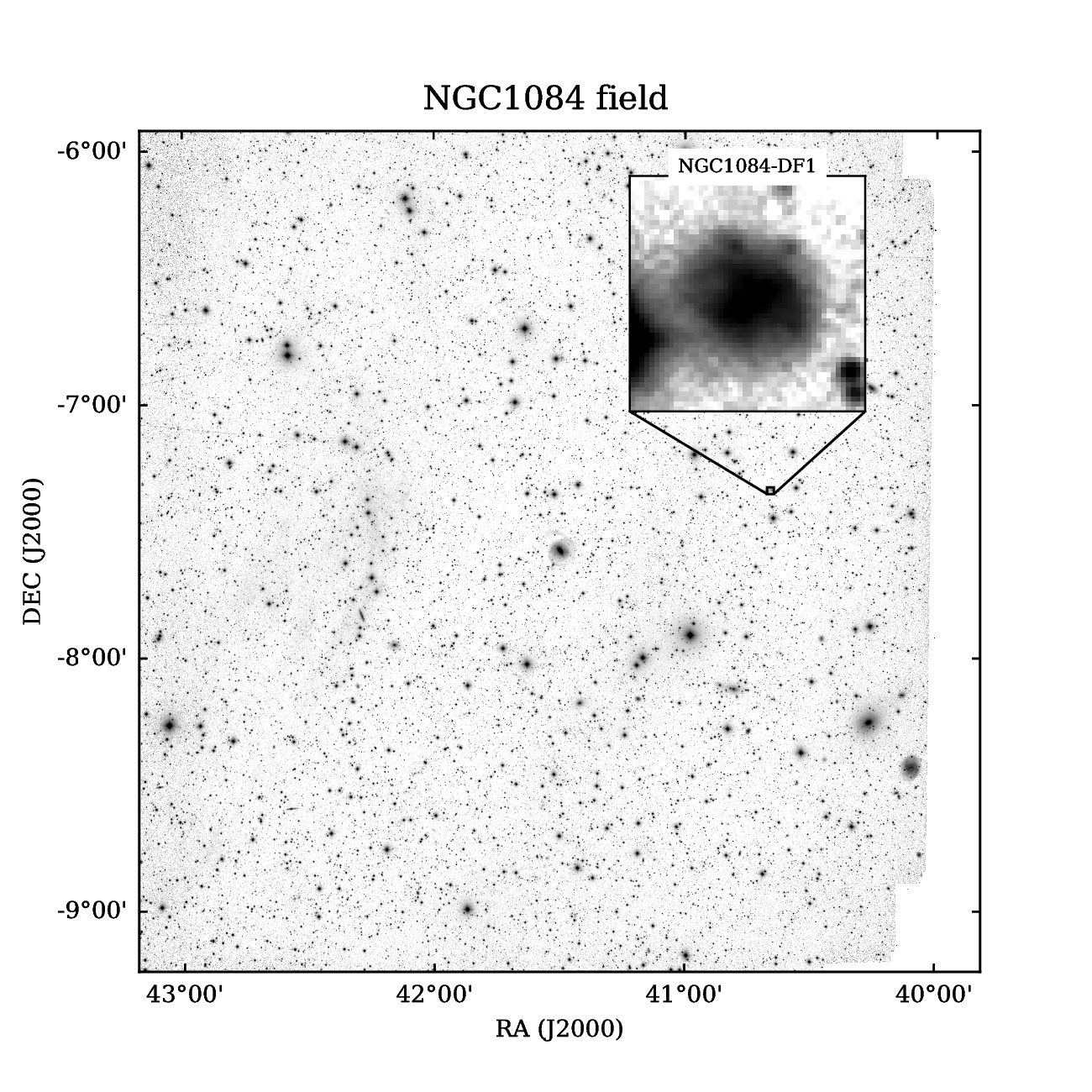}{0.5\textwidth}{}}
\caption{Deep Dragonfly images (stacked \textit{g}+\textit{r}) of the fields centered on NGC\,4258 (top left), NGC\,3384 (in the M\,96 group) (top right), NGC\,1052 (bottom left), and NGC\,1084 (bottom right).
The zoomed insets show the LSB galaxies and span $2'$ on a side.
Within each of the four panels, the pixel stretch of the main frame is the same as that of all the zoomed insets, and the zoom is the same for each of the insets.
All frames have North up and East to the left.
Each Dragonfly frame spans approximately $3\degree \times 3\degree$. \label{fig:fields}}
\end{figure*}

The LSB candidates were identified by visually inspecting the Dragonfly images.
This method is not objective and thus the sample is likely to be incomplete;
in the future we will re-analyze the Dragonfly data using more objective techniques.
The 23 candidates which were subsequently observed with \textit{HST} are shown in the Dragonfly images of Figure \ref{fig:fields}.
Their basic information can be found in Tables \ref{tab:table1} and \ref{tab:gftable}.
Sixteen of the galaxies had been seen before, in visual inspections of deep photographic plates and other more recent surveys.
The designations and references given to these previous detections are given in Table \ref{tab:table1}.
The remaining seven are new discoveries; they are typically the fainter objects in the sample.
Note that all these galaxies are bright, unambiguous detections (see Figure \ref{fig:fields}); we are not probing the limits of the Dragonfly imaging in this paper.

\begin{deluxetable*}{lllll}[]
\tabletypesize{\scriptsize}
\tablecaption{Designations and Positions \label{tab:table1}}
\tablenum{1}
\tablehead{\colhead{Id} & \colhead{alt. Id} & \colhead{ref} & \colhead{RA (J2000)} & \colhead{DEC (J2000)} }
\startdata
NGC\,1084-DF1  & WHI B0240-07 & \cite{WHI}       & 02:42:38.0  & $-$07:20:16.3  \\
NGC\,1052-DF4  & new        &       --       & 02:39:15.1  & $-$08:06:58.6   \\
NGC\,1052-DF2  & {[}KKS2000{]}04   & \cite{kks2000}    & 02:41:46.8  & $-$08:24:09.3  \\
NGC\,4258-DF1  & {[}KK98{]}136     & \cite{kk1998}       & 12:20:40.2 & $+$47:00:02.7    \\
M\,96-DF11 & {[}KK98{]}093    & \cite{kk1998}         & 10:46:24.6 & $+$14:01:27.8    \\
NGC\,4258-DF2  & {[}KKH2011{]}S09 & \cite{kkh2011}  & 12:19:36.0 & $+$47:05:35.3   \\
NGC\,4258-DF3  & BTS 109 & \cite{BTS} & 12:17:09.4  & $+$47:03:49.2   \\
M\,96-DF9  & {[}KK2004{]} LeG 14    & \cite{kk2004}               & 10:46:14.2 & $+$12:57:38.0   \\
M\,96-DF1  & new             & --         & 10:48:13.1 & $+$11:58:06.4   \\
M\,96-DF8  & {[}KK98{]}094  & \cite{kk1998}          & 10:46:57.4 & $+$12:59:55.3   \\
NGC\,1052-DF8  & new            & --          & 02:40:45.5  & $-$07:38:48.7   \\
M\,96-DF4  & {[}KK98{]}096     & \cite{kk1998}      & 10:50:27.2 & $+$12:21:35.3   \\
M\,96-DF3  & {[}KK2004{]} LeG 9  & \cite{kk2004}                 & 10:42:34.4 & $+$12:09:04.4    \\
NGC\,4258-DF6  & {[}KK98{]}132    & \cite{kk1998}       & 12:19:06.2  & $+$47:43:49.3   \\
NGC\,1052-DF5  & new   & --                   & 02:39:12.6  & $-$08:08:27.3   \\
M\,96-DF5  & dw1049+12b         & \cite{muller18}       & 10:49:26.0 & $+$12:33:10.3   \\
M\,96-DF7  & Leo I GROUP:{[}TT2002{]} 10   & \cite{TT2002}                   & 10:47:13.5 & $+$12:48:09.1   \\
NGC\,4258-DF5  & LV J1215+4732 & \cite{kkm2014}            & 12:15:51.4 & $+$47:32:51.5   \\
NGC\,1052-DF7  & new  & --                      & 02:38:29.9  & $-$07:55:33.8   \\
M\,96-DF10 & dw1048+13 & \cite{muller18}                      & 10:48:36.0 & $+$13:03:35.2  \\
M\,96-DF6  & Leo I GROUP:{[}TT2002{]} 09         & \cite{TT2002}             & 10:46:53.1 & $+$12:44:33.5   \\
NGC\,1052-DF1  & new     & --                 & 02:40:04.6   & $-$08:26:44.4  \\
M\,96-DF2  & new            & --          & 10:47:40.6 & $+$12:02:55.8  \\
\enddata
\tablecomments{Objects sorted by effective surface brightness, as in Figures \ref{fig:lineuprgb} and \ref{fig:lineupgrey}. }
\end{deluxetable*}

\subsection{HST Imaging} \label{sec:HST imaging}

\begin{figure*}[]
\epsscale{1.0}
\plotone{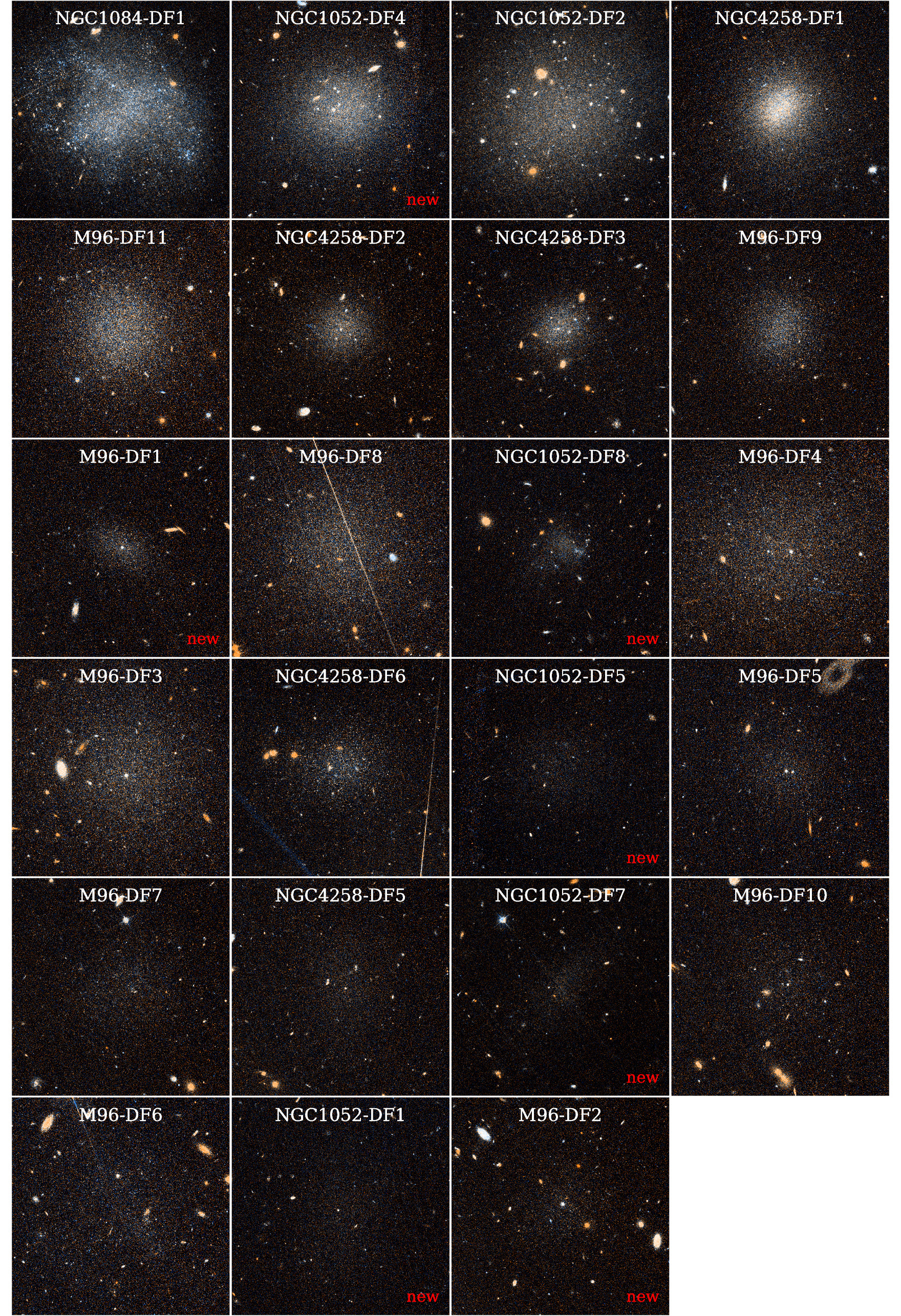}
\caption{Pseudo-color images created from \textit{HST}/ACS $V_{606}$ and $I_{814}$ images using the \cite{lupton} algorithm. From left to right and top to bottom, the galaxies are ordered by effective surface brightness in the $V_{606}$-band. All panels are displayed using the same pixel stretch. Each panel spans $1'$ on a side. North is up and East is to the left. Newly discovered galaxies are marked with `new' in the corner.
\label{fig:lineuprgb}}
\end{figure*}

\begin{figure*}[]
\epsscale{1.0}
\plotone{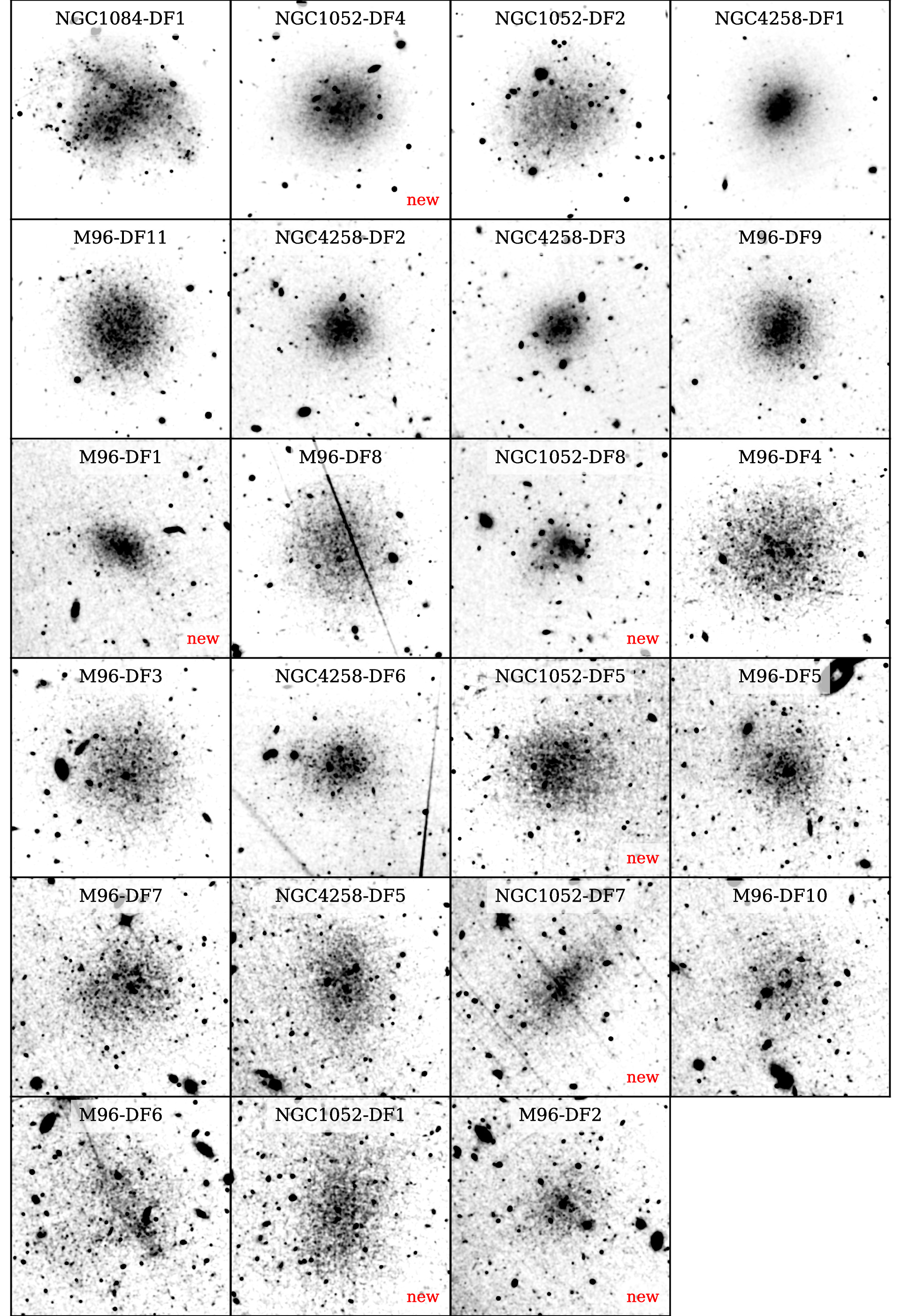}
\caption{Similar to Figure \ref{fig:lineuprgb} but showing the stacked $V_{606}+I_{814}$ frames after smoothing with a gaussian filter. Unlike in Figure \ref{fig:lineuprgb}, each panel is displayed using a different pixel stretch that makes the respective galaxy easily visible. Newly discovered galaxies are marked with `new' in the corner.
\label{fig:lineupgrey}}
\end{figure*}

As part of our Cycle 24 \textit{HST} program 14644, we observed 23 candidate LSBs with the ACS/WFC.
In the NGC\,4258 and M\,96 fields we obtained a single orbit per galaxy, split into two dithered exposures in the $I_{814}$ filter totaling 1164 (NGC\,4258) and 1096 (M\,96) seconds, and two dithered exposures in the $V_{606}$ filter totaling 1094 (NGC\,4258) and 1026 (M\,96) seconds.
In the NGC\,1052 and NGC\,1084 fields we obtained two orbits per galaxy, split into four dithered exposures in the $I_{814}$ filter totaling 2320 seconds and four dithered exposures in the $V_{606}$ filter totaling 2180 seconds.
The choices of these exposure times were motivated by the distances of the primary galaxies in these fields and, in turn, the estimated depth needed to determine distances of the low surface brightness objects (Section \ref{sec:trgb}) if they are associated with the primary galaxies.
For reference, we adopt the following distances:
For M\,96 we adopt the group distance of \cite{Tully:2013}, $D = 10.7 \pm 0.3$\,Mpc, which is derived from a combination of Cepheid, SBF, and Tully-Fisher measurements.
For NGC\,4258 we use the highly accurate maser distance, $D = 7.6 \pm 0.2$\,Mpc \citep{Humphreys:2013}.
For NGC\,1052 we adopt $D = 20 \pm 1$\,Mpc which is an average of SBF measurements from \cite{Blakeslee:2001} and \cite{Tully:2013}.
For NGC\,1084 we adopt $D = 18.5 \pm 1$\,Mpc which is rough average from recent supernova \citep{Barbarino:2015} and Tully-Fisher \citep{Tully:2016} measurements.

The ACS exposures were bias- and dark-current-subtracted, flatfielded, and CTE-corrected by the standard STScI pipeline, resulting in calibrated `flc' files.
Using the latest version of DrizzlePac\footnote{http://www.stsci.edu/hst/HST\_overview/drizzlepac}, we produced our own drizzled frames from the `flc' files after cleaning them from cosmic rays using LA-Cosmic \citep{lacosmic} and aligning them using the TWEAKREG utility of DrizzlePac.
All galaxies were drizzled to a reference frame that has North up and East to the left, with pixel size of $0.05''$.
Pseudo-color images of all the objects in the sample, generated from the drizzled frames, ordered by effective surface brightness, are shown in Figure \ref{fig:lineuprgb}.
In this figure, all panels are shown using the same pixel stretch in order to enable meaningful comparison in surface brightness levels and overall color.
However, it is difficult to see some of the most diffuse objects on this stretch; we therefore also include Figure \ref{fig:lineupgrey} which displays the $V_{606}+I_{814}$ data after smoothing and using a different pixel stretch in each panel appropriate for the surface brightness of the particular galaxy.

Visually, most of the galaxies appear to be featureless ``blobs'', superficially resembling dwarf spheroidal galaxies in the Local Group.
We see no galaxies with spiral arms or other regular morphological substructure.
Against the backdrop of this overall similarity, we do see differences within the sample including:
\begin{itemize}
\item Several of the galaxies appear to contain one or more star cluster-like objects and/or a bright, central nuclear star cluster, while others have none.
\item One of the galaxies, NGC\,1052-DF2, was recently identified as a dark matter deficient galaxy from kinematic measurements of its globular clusters \citep{df2nature,df2rn}; the clusters themselves are also interesting as their average brightness and degree of flattening are unusually high \citep{df2apj}.
\item Blue, star-forming clumps are clearly visible in NGC\,1084-DF1 and NGC\,1052-DF8 (and both objects are detected, at least visually, in archival \textit{GALEX} observations), while most of the others appear much smoother and redder.
\end{itemize}

In what follows, we use the ACS imaging data to quantify the structure of the galaxies and derive distance estimates.
We use AB magnitudes everywhere except for the stellar photometry and TRGB analysis (Section \ref{sec:trgb}).

\section{Structure and Brightness}
We used GALFIT \citep{galfit} to measure the structure, magnitude, color, and surface brightness of the galaxies.
The fitting procedure involved several steps, which we summarize here.
In order to ensure stable fits to these faint galaxies, we aggressively masked all contaminating sources.
To do this, we first created an initial mask for each frame by running SExtractor \citep{sextractor} on the drizzled $I_{814}$ frames.
SExtractor was run with a relatively high threshold to avoid detecting the low surface brightness galaxies themselves; the goal is to detect only the contaminating sources.
In order to improve the masking of the diffuse light around the edges of bright sources, we then grew the mask by smoothing it with a Gaussian and absorbing all pixels higher than some threshold value (typically 0.05, where 1 is the value of a masked pixel in the original mask and 0 is the value of an unmasked pixel) into the mask.
We also added some features by hand, in particular diffraction spikes from bright stars and other artifacts.
We also make sure that the nuclear and globular star clusters assumed to be associated with the LSB galaxies (Section \ref{sec:starclusters}) are masked.

GALFIT was run on the summed, drizzled $I_{814} + V_{606}$ images.
We used the summed images as they provide more stable fits than the (lower S/N) individual bands. 
A single S\'{e}rsic function was used as the model, allowing all parameters to vary.
For each galaxy, we ran the fits multiple times, each time updating the fitting region, the `guess values' for the model parameters, and the mask, until the model parameters converged and the residuals were deemed satisfactory.
The final model fits and residuals are shown in Figure \ref{fig:galfitgridspec}.
In order to measure the $V_{606}$ and $I_{814}$ magnitudes, we then ran GALFIT on the $V_{606}$ and $I_{814}$ frames separately, holding all parameters (except for the magnitude) fixed to the values of the best fit model determined for the summed image.
From the model parameters, we calculated the central and effective surface brightness (defined as the average surface brightness within the effective radius) for each galaxy using equations provided in \cite{sersic}.
Measured properties are given in Table \ref{tab:gftable}.

\begin{figure}[]
\epsscale{1.15}
\plotone{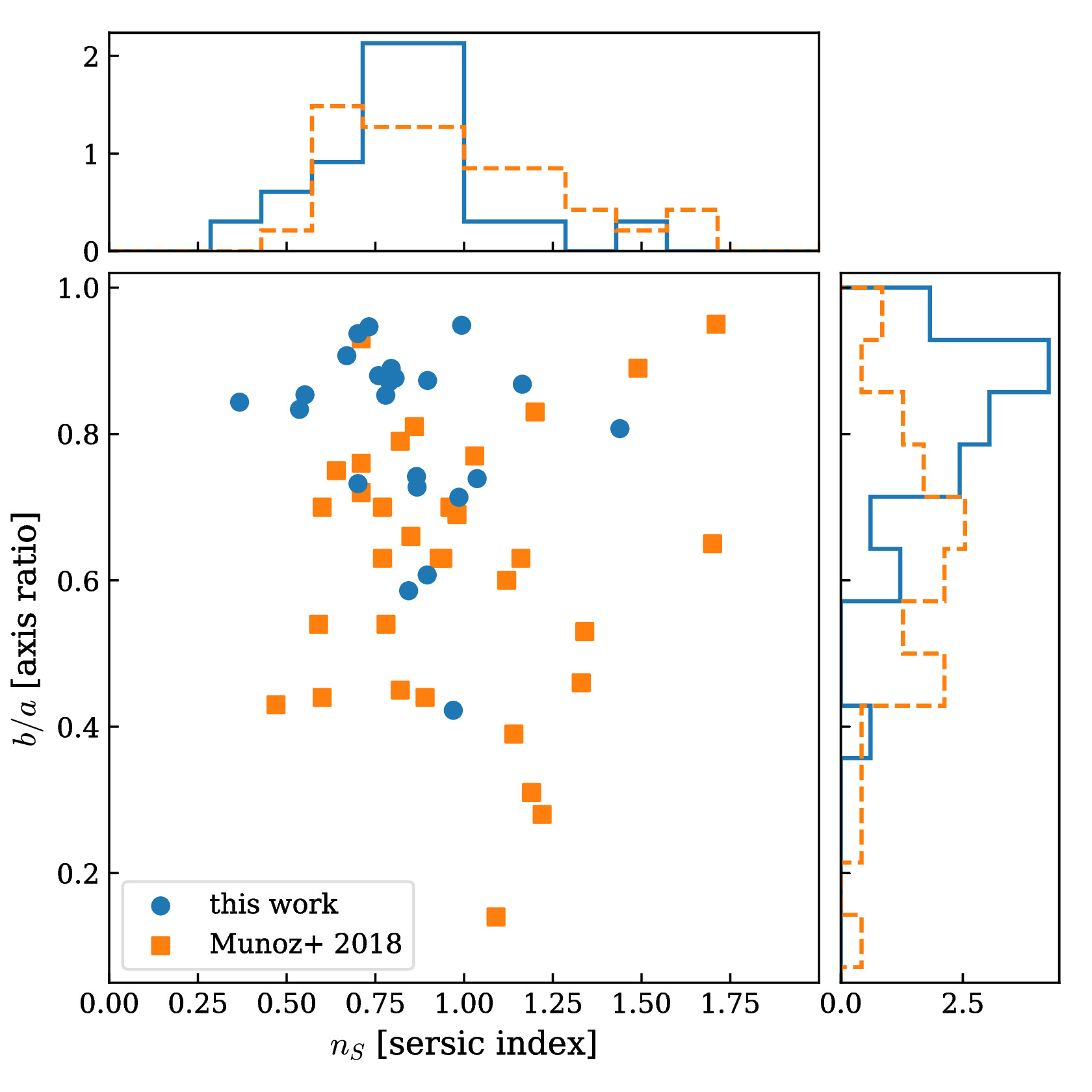}
\caption{S\'{e}rsic indices and axis ratios of the galaxies in this sample (blue circles and blue, solid histograms), measured using GALFIT.
It is seen that the majority of these objects are fairly round ($\langle b/a \rangle = 0.85$) and have light profiles slightly less concentrated than an exponential ($\langle n_S \rangle = 0.8$).
Also plotted are the outer halo satellites of the Milky Way (orange squares and orange, dashed histograms), from Table 5 of \cite{munoz18}.
Only objects with $R_e > 25$\,pc from their Table 5 are plotted, in order to remove globular clusters.
\label{fig:ba_vs_sersic}}
\end{figure}

\begin{figure*}[]
\plotone{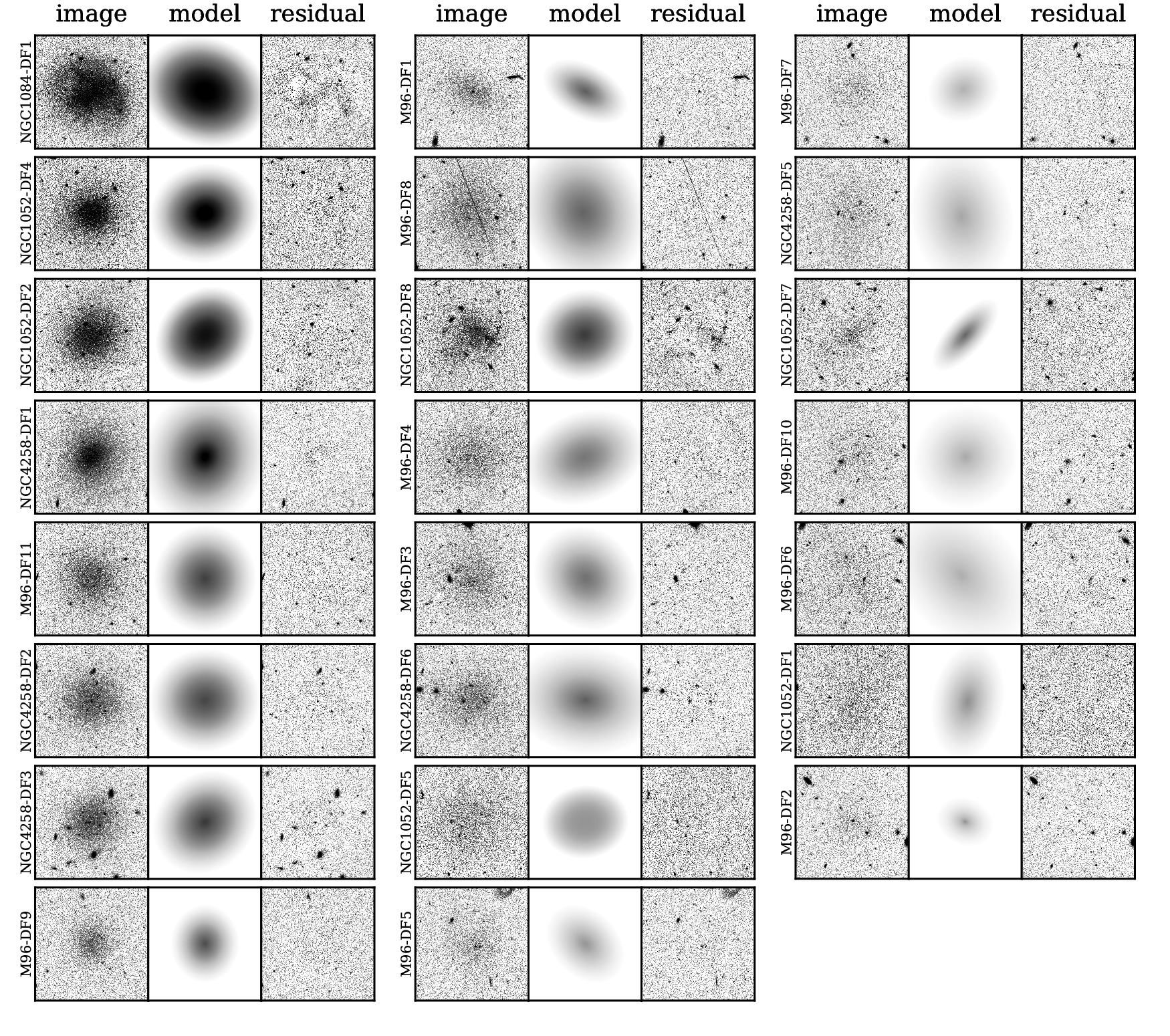}
\caption{GALFIT models for all the galaxies in the sample, ordered by effective surface brightness. Shown for each object is the fitting region in the stacked $V+I$ image (left), best fitting GALFIT model (middle), and corresponding residuals (right). All panels are displayed with the same pixel stretch. Note that the size of the fitting region is not the same for each galaxy; the boxes do not all have the same size. North is up and East is to the left. \label{fig:galfitgridspec}}
\end{figure*}

As expected, the galaxies are very diffuse, having a median $V_{606}$ central surface brightness $\langle \mu_{0,V} \rangle = 24.7 \ \rm mag \ arcsec^{-2}$ and median effective surface brightness  $\langle \mu_{e,V} \rangle = 26.0 \ \rm mag \ arcsec^{-2}$.
In Figure \ref{fig:ba_vs_sersic} we show the distribution of S\'{e}rsic indices and axis ratios for the sample; there is no significant relation.
The majority of the galaxies are well-fitted by a S\'{e}rsic function of S\'{e}rsic index $n < 1$, with none exceeding $n \sim 1.5$ and a median $\langle n_S \rangle = 0.8$.
Barring two or three clear exceptions, most of the galaxies are remarkably round, with a median axis ratio $\langle b/a \rangle = 0.85$.
The immediate implication is that the galaxies are not a population of thin disks, as they would then have an average axis ratio of $\sim 0.5$.

\begin{deluxetable*}{lllllllll}[]
\tabletypesize{\scriptsize}
\tablecaption{Observed Properties \label{tab:gftable}}
\tablenum{2}
\tablehead{\colhead{Id} & \colhead{${V_{\rm 606}}$\,\tablenotemark{a}} & \colhead{${\mu_{0,V}}$\,\tablenotemark{b}} & \colhead{${\mu_{e,V}}$\,\tablenotemark{c}} & \colhead{$V_{\rm 606}-I_{\rm 814}$} & \colhead{${r_{\rm eff}}$\,\tablenotemark{d}} & \colhead{$n$\,\tablenotemark{e}} & \colhead{$b/a$\,\tablenotemark{f}} }
\startdata
NGC\,1084-DF1  & 16.2 $\pm$ 0.1 & 23.9 $\pm$ 0.1 & 24.7 $\pm$ 0.1 & 0.19 $\pm$ 0.1 & 18.2 $\pm$ 0.1 & 0.54 $\pm$ 0.01 & 0.83 $\pm$ 0.01    \\
NGC\,1052-DF4  & 16.5 $\pm$ 0.1 & 23.7 $\pm$ 0.1 & 25.1 $\pm$ 0.1 & 0.32 $\pm$ 0.1 & 16.5 $\pm$ 0.1 & 0.79 $\pm$ 0.01 & 0.89 $\pm$ 0.01     \\
NGC\,1052-DF2  & 16.2 $\pm$ 0.1 & 24.2 $\pm$ 0.1 & 25.1 $\pm$ 0.1 & 0.40 $\pm$ 0.1 & 21.2 $\pm$ 0.1 & 0.55 $\pm$ 0.01 & 0.85 $\pm$ 0.01     \\
NGC\,4258-DF1  & 16.7 $\pm$ 0.1 & 23.0 $\pm$ 0.1 & 25.2 $\pm$ 0.1 & 0.44 $\pm$ 0.1 & 15.2 $\pm$ 0.1 & 1.16 $\pm$ 0.01 & 0.87 $\pm$ 0.01     \\
M\,96-DF11     & 16.8 $\pm$ 0.1 & 24.0 $\pm$ 0.1 & 25.2 $\pm$ 0.1 & 0.45 $\pm$ 0.1 & 15.7 $\pm$ 0.1 & 0.73 $\pm$ 0.01 & 0.95 $\pm$ 0.01     \\
NGC\,4258-DF2  & 18.2 $\pm$ 0.1 & 24.1 $\pm$ 0.1 & 25.3 $\pm$ 0.1 & 0.49 $\pm$ 0.1 & 8.4 $\pm$ 0.1  & 0.70 $\pm$ 0.01 & 0.94 $\pm$ 0.01     \\
NGC\,4258-DF3  & 18.2 $\pm$ 0.1 & 23.9 $\pm$ 0.1 & 25.3 $\pm$ 0.1 & 0.45 $\pm$ 0.1 & 8.6 $\pm$ 0.1  & 0.79 $\pm$ 0.01 & 0.87 $\pm$ 0.01     \\
M\,96-DF9      & 17.8 $\pm$ 0.1 & 24.1 $\pm$ 0.1 & 25.7 $\pm$ 0.1 & 0.39 $\pm$ 0.1 & 12.1 $\pm$ 0.1 & 0.90 $\pm$ 0.01 & 0.87 $\pm$ 0.01     \\
M\,96-DF1      & 18.8 $\pm$ 0.1 & 24.3 $\pm$ 0.2 & 25.8 $\pm$ 0.2 & 0.37 $\pm$ 0.1 & 9.6 $\pm$ 0.1  & 0.84 $\pm$ 0.01 & 0.59 $\pm$ 0.01      \\
M\,96-DF8      & 16.6 $\pm$ 0.1 & 24.5 $\pm$ 0.1 & 25.8 $\pm$ 0.1 & 0.55 $\pm$ 0.1 & 23.4 $\pm$ 0.1 & 0.78 $\pm$ 0.01 & 0.85 $\pm$ 0.01     \\
NGC\,1052-DF8  & 19.3 $\pm$ 0.1 & 24.7 $\pm$ 0.2 & 25.8 $\pm$ 0.2 & 0.40 $\pm$ 0.1 & 6.7 $\pm$ 0.1  & 0.67 $\pm$ 0.01 & 0.91 $\pm$ 0.01     \\
M\,96-DF4      & 16.5 $\pm$ 0.1 & 24.7 $\pm$ 0.1 & 25.9 $\pm$ 0.1 & 0.46 $\pm$ 0.1 & 27.4 $\pm$ 0.1 & 0.70 $\pm$ 0.01 & 0.73 $\pm$ 0.01     \\
M\,96-DF3      & 16.5 $\pm$ 0.1 & 24.6 $\pm$ 0.1 & 25.9 $\pm$ 0.1 & 0.42 $\pm$ 0.1 & 24.7 $\pm$ 0.1 & 0.76 $\pm$ 0.01 & 0.88 $\pm$ 0.01     \\
NGC\,4258-DF6  & 17.5 $\pm$ 0.1 & 24.3 $\pm$ 0.3 & 26.2 $\pm$ 0.3 & 0.25 $\pm$ 0.1 & 18.5 $\pm$ 0.2 & 1.04 $\pm$ 0.01 & 0.74 $\pm$ 0.01     \\
NGC\,1052-DF5  & 19.3 $\pm$ 0.1 & 25.8 $\pm$ 0.2 & 26.3 $\pm$ 0.2 & 0.21 $\pm$ 0.1 & 9.2 $\pm$ 0.1  & 0.37 $\pm$ 0.01 & 0.84 $\pm$ 0.01     \\
M\,96-DF5      & 18.2 $\pm$ 0.1 & 25.0 $\pm$ 0.3 & 26.6 $\pm$ 0.3 & 0.44 $\pm$ 0.1 & 16.3 $\pm$ 0.2 & 0.87 $\pm$ 0.01 & 0.74 $\pm$ 0.01      \\
M\,96-DF7      & 17.9 $\pm$ 0.1 & 25.2 $\pm$ 0.3 & 26.6 $\pm$ 0.3 & 0.08 $\pm$ 0.1 & 17.6 $\pm$ 0.2 & 0.81 $\pm$ 0.01 & 0.88 $\pm$ 0.01     \\
NGC\,4258-DF5  & 17.7 $\pm$ 0.1 & 25.3 $\pm$ 0.5 & 26.8 $\pm$ 0.6 & 0.37 $\pm$ 0.1 & 23.0 $\pm$ 0.5 & 0.87 $\pm$ 0.02 & 0.73 $\pm$ 0.01     \\
NGC\,1052-DF7  & 18.9 $\pm$ 0.1 & 25.1 $\pm$ 0.3 & 26.9 $\pm$ 0.3 & 0.29 $\pm$ 0.1 & 17.7 $\pm$ 0.2 & 0.97 $\pm$ 0.01 & 0.42 $\pm$ 0.01     \\
M\,96-DF10     & 17.8 $\pm$ 0.1 & 25.2 $\pm$ 0.7 & 27.0 $\pm$ 0.7 & 0.27 $\pm$ 0.1 & 20.4 $\pm$ 0.6 & 0.99 $\pm$ 0.03 & 0.95 $\pm$ 0.01     \\
M\,96-DF6      & 16.6 $\pm$ 0.1 & 25.2 $\pm$ 0.9 & 27.0 $\pm$ 0.9 & 0.19 $\pm$ 0.1 & 40.6 $\pm$ 1.4 & 0.99 $\pm$ 0.03 & 0.71 $\pm$ 0.01     \\
NGC\,1052-DF1  & 18.2 $\pm$ 0.1 & 25.8 $\pm$ 0.4 & 27.4 $\pm$ 0.5 & 0.31 $\pm$ 0.1 & 25.9 $\pm$ 0.4 & 0.90 $\pm$ 0.01 & 0.61 $\pm$ 0.01     \\
M\,96-DF2      & 18.3 $\pm$ 0.1 & 24.8 $\pm$ 0.7 & 27.6 $\pm$ 0.8 & 0.26 $\pm$ 0.1 & 21.6 $\pm$ 0.6 & 1.44 $\pm$ 0.03 & 0.81 $\pm$ 0.01    \\
\enddata
\tablecomments{Objects sorted by effective surface brightness, as in Figures \ref{fig:lineuprgb} and \ref{fig:lineupgrey}. Uncertainties represent the random uncertainties in the model parameter fits as reported by GALFIT.}
\tablenotetext{a}{Integrated apparent magnitude} \vspace{-0.25cm}
\tablenotetext{b}{Central surface brightness ($\rm mag \, arcsec^{-2}$)} \vspace{-0.25cm}
\tablenotetext{c}{Effective surface brightness ($\rm mag \, arcsec^{-2}$)} \vspace{-0.25cm}
\tablenotetext{d}{Semi-major axis effective radius (arcsec)} \vspace{-0.25cm}
\tablenotetext{e}{S\'{e}rsic index} \vspace{-0.25cm}
\tablenotetext{f}{Axis ratio} \vspace{-0.25cm}
\end{deluxetable*}

\section{Stellar Photometry and TRGB distances \label{sec:trgb}}
Inspection of the high-resolution ACS images reveals that many (although not all) of the galaxies are clearly resolved into stars.
To measure the stellar photometry, we used the publicly available ACS module of DOLPHOT, a modified version of HSTphot \citep{dolphot}.
The photometry was carried out on the calibrated `flc' files for all galaxies in the sample, even those that did not appear to be obviously resolved into stars.
Artificial star tests were also performed on the images, using the artificial star utilities in DOLPHOT, in order to quantify the photometric errors and completeness.
We followed the usual procedures for performing ACS photometry as outlined in the manual\footnote{http://americano.dolphinsim.com/dolphot/dolphotACS.pdf}, including pre-processing steps such as hot pixel masking, sky determination, and image alignment.
DOLPHOT parameters were set as recommended in the manual for these kind of data.

The final catalogs produced by DOLPHOT include the position, flux, flux uncertainty, sharpness, crowding, $\chi^2$, and many other parameters for each photometered source.
As in previous studies \citep[e.g.][]{ANGST,GHOSTS,danieli2016,mcquinn}, we apply strict cuts to the output catalogs in order to produce high quality final catalogs.
We adopt the crowding and sharpness cuts used by \cite{GHOSTS}, and the signal to noise ratio cuts used by \cite{mcquinn}.

\begin{figure*}[]
\epsscale{1}
\plotone{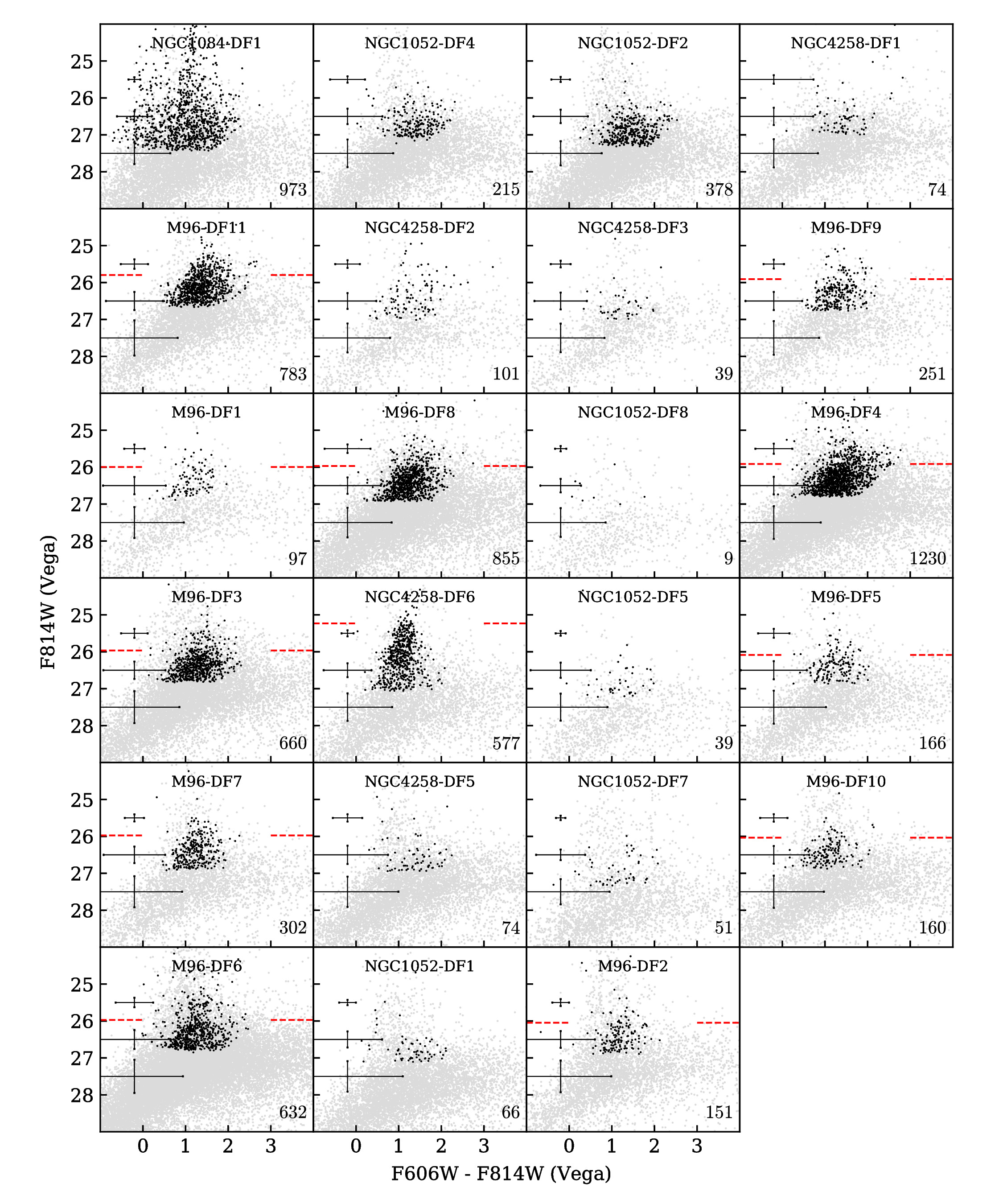}
\caption{Color--magnitude diagrams of galaxies in our sample, imaged with \textit{HST}/ACS and photometered with DOLPHOT.
The galaxies are ordered by effective surface brightness as in Figs \ref{fig:lineuprgb} and \ref{fig:lineupgrey}.
Light gray points show all detected point sources before applying quality cuts to the photometry.
Black points show detections remaining after applying quality cuts.
The number in the bottom right indicates the number of black points.
Error bars on the left indicate typical photometric uncertainties in bins of ${I_{\rm 814}}$.
The red dashed line indicates the measured TRGB magnitude, where applicable.
\label{fig:cmds}}
\end{figure*}

Figure \ref{fig:cmds} presents the color--magnitude diagrams (CMDs) for the sample.
For each galaxy, the CMD is populated by the detections enclosed within an elliptical region of size $2 R_e$ centered on the galaxy, where $R_e$, as well as the axis ratio and position angle of the ellipse, correspond to the best fitting S\'{e}rsic model from GALFIT.
The light grey points show all detections before applying the quality cuts, while the black points show the detections remaining after applying the cuts.
The number in the bottom right of each panel indicates the number of black points.
The error bars on the left indicate the mean photometric uncertainties, in bins of ${I_{\rm 814}}$, as measured by DOLPHOT.
The photometry has also been corrected for Galactic extinction using NED's Galactic Extinction Calculator\footnote{https://ned.ipac.caltech.edu/forms/calculator.html}.
As we can see, several of the galaxies are almost completely unresolved into stars while others exhibit a clearly identifiable red giant branch (RGB) (see Figure 6 of \cite{GHOSTS} for an illustration of the various features present in color magnitude diagrams).

In order to measure distances from CMDs, we use the tip of the red giant branch (TRGB) method.
One of the most efficient and precise approaches to measuring distances in the local universe, the TRGB method is a standard--candle distance indicator based on the predictable maximum luminosity of red giant branch (RGB) stars, $M_{\rm TRGB}^{I} \approx -4.05$, just prior to the helium flash\footnote{Note that the ``tip'' of the RGB is not a limit, but the magnitude above which the slope of the luminosity function steepens. In well-populated CMDs there are many giants that are brighter than the tip, including all stars on the AGB.} \citep{mcquinn}.
To make a quantitative measurement of the TRGB, which manifests as a discontinuity or `edge' in the luminosity function, we used the logarithmic edge-detection filter described in \cite{mendez} applied to a Gaussian-smoothed LF as in \cite{seth}.
The stars used in the measurements are the black points in the CMDs in Figure \ref{fig:cmds}.

\begin{figure}[]
\epsscale{1.2}
\plotone{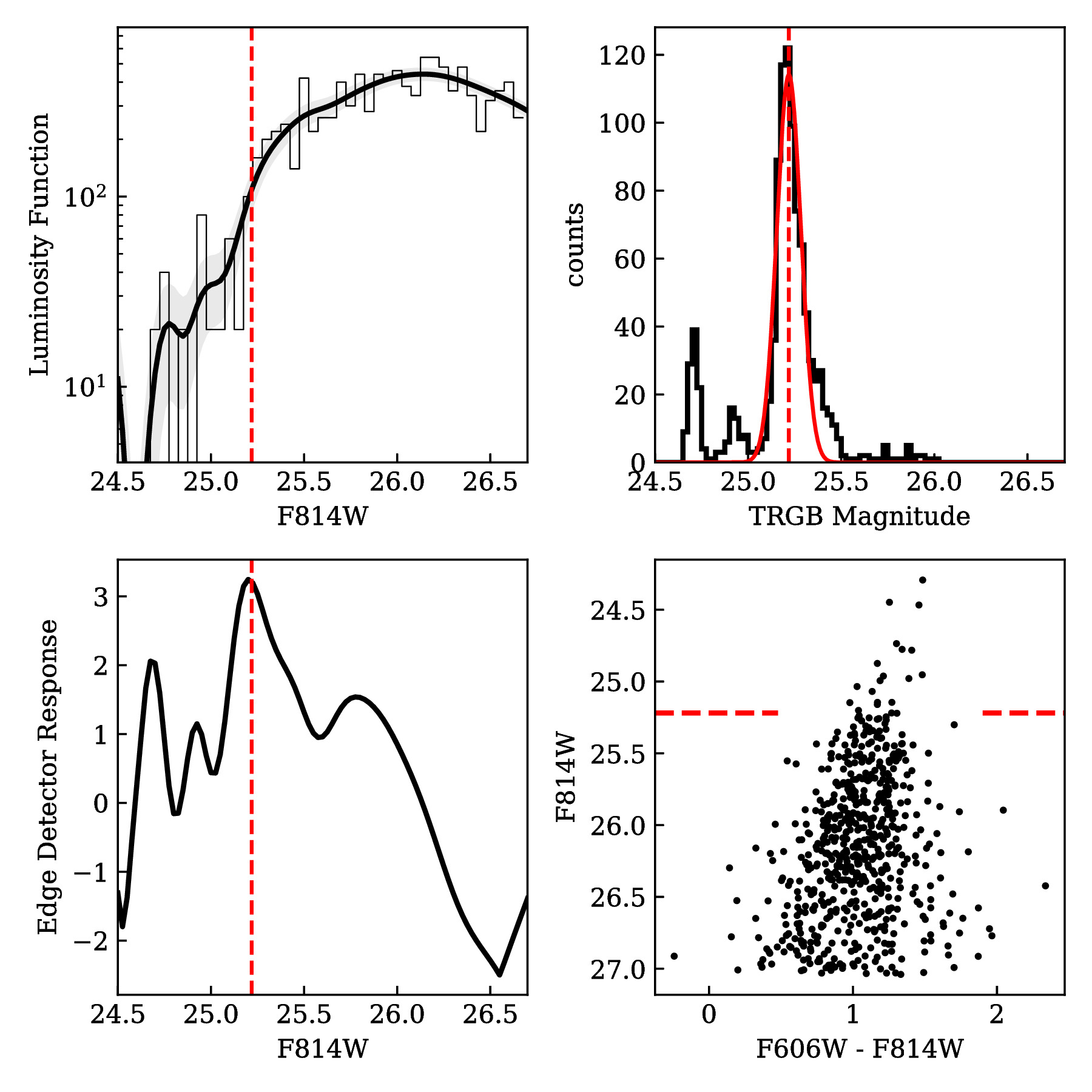}
\caption{TRGB fitting for NGC\,4258-DF6, showing the $I_{814}$ luminosity function (top left, step-plot), Gaussian-smoothed luminosity function and $1\sigma$ uncertainty regions from bootstrap resamples (top left, solid line and shaded region), edge-detector response (bottom left), distribution of TRGB magnitudes from Monte Carlo trials (top right), and the color--magnitude diagram of stars used in the measurement (bottom right). The measured TRGB magnitude is shown as the dotted line in each panel. \label{fig:trgb}}
\end{figure}

The TRGB magnitudes and associated uncertainties were determined by running 1000 Monte Carlo trials with bootstrap resampling of the stars.
In each trial, Gaussian random errors based on artificial star tests were added to the resampled photometry.
Each trial returned the magnitude corresponding to the peak of the weighted edge-detector response function.
We then generated a histogram of the resulting magnitudes and fitted the peak with a Gaussian.
We take the mean and spread of this Gaussian to be the magnitude of the TRGB and its random uncertainty.
As in \cite{mendez}, the edge-detector response was weighted by the Poisson noise in the smoothed LF in order to reduce the effect of noise spikes generated by large variations in the LF at magnitudes brighter than the TRGB due to the small number of asymptotic giant branch (AGB) stars.
However, in some Monte Carlo trials, there were still additional peaks in the edge-detector response that clearly do not correspond to the TRGB magnitude.
In these cases, we initialized the Gaussian fit close to the likely TRGB peak.
In all cases, we scrutinized the results by examining diagnostic plots like the ones shown in Figure \ref{fig:trgb}.
We find that this method provides well-converged estimates for the galaxies with a well-populated RGB above the photometric limit.
We also verified that the  maximum likelihood technique of \cite{makarov} gives similar results to the edge detection method.

The observed TRGB magnitudes can be converted to distances using the absolute magnitude of the TRGB.
The absolute magnitude of the TRGB in the $I_{814}$ filter has a well-known slight dependence on metallicity, which can be taken into account by using a color-dependent correction.
We use the calibration from \cite{rizzi} derived for the \textit{HST}/ACS filters, reproduced here:
\begin{equation*}
M^{I_{814}}_{\rm ACS} = -4.06 + 0.2 \cdot [(V_{606} - I_{814}) - 1.23].
\end{equation*}
Distance estimates were then calculated by drawing 1000 magnitude values from the Gaussian used to fit the TRGB magnitude, and assuming the above absolute magnitude calibration, where the $V_{606}-I_{814}$ color used corresponds to the median color of the stars at the measured TRGB magnitude.
We take the mean and standard deviation of the resulting distribution to be the distance estimate and its random uncertainty.

In order to apply this method, the RGB of the CMD needs to be sufficiently populated well above the photometric limit of the observations.
This is the case for all galaxies in the M\,96 field as well as NGC\,4258-DF6; our measurement of their TRGB magnitudes are indicated as red, dashed lines in Figure \ref{fig:cmds} and corresponding distances are shown as the blue points and uncertainties in Figure \ref{fig:distcompare}.
Although we attempted to measure it, we do not identify the TRGB in the other galaxies in our sample for one or more of the following reasons: (1) the CMDs are too sparsely populated (as in NGC\,4258-DF2 and NGC\,4258-DF3), (2) there is too much contamination from bright stars (as in the star forming galaxy NGC\,1084-DF1), (3) there is a concentration of detections near the photometric limit, but it is caused by noise peaks and blending (as in NGC\,1052-DF2 and NGC\,1052-DF4), as we verified using artificial star simulations \citep{vd:2018d}.
The absence of a TRGB well above the photometric limit immediately places lower limits on the distances to those objects.
As in \cite{kara2017}, assuming that the TRGB has absolute magnitude $M_{\rm TRGB}^{I_{814}} = -4.05$ and $V_{606}-I_{814} \approx 1$, we can obtain a very stringent lower limit by asserting that the TRGB cannot be brighter than the brightest stars in the CMD with $V_{606}-I_{814} \approx 1$.
The lower limit distances shown in Figure \ref{fig:distcompare} were derived using this logic.
Note that these limits are very conservative for CMDs with many detected objects; furthermore, some of the CMDs may have noise/contamination detections at bright magnitudes which falsely drive the limits to lower values.
In the case of NGC\,1084-DF1, the limit is driven down by bright, non-RGB stars with $V_{606}-I_{814} \sim 1$.

\section{SBF Distances}

\begin{figure}[]
\epsscale{1.2}
\plotone{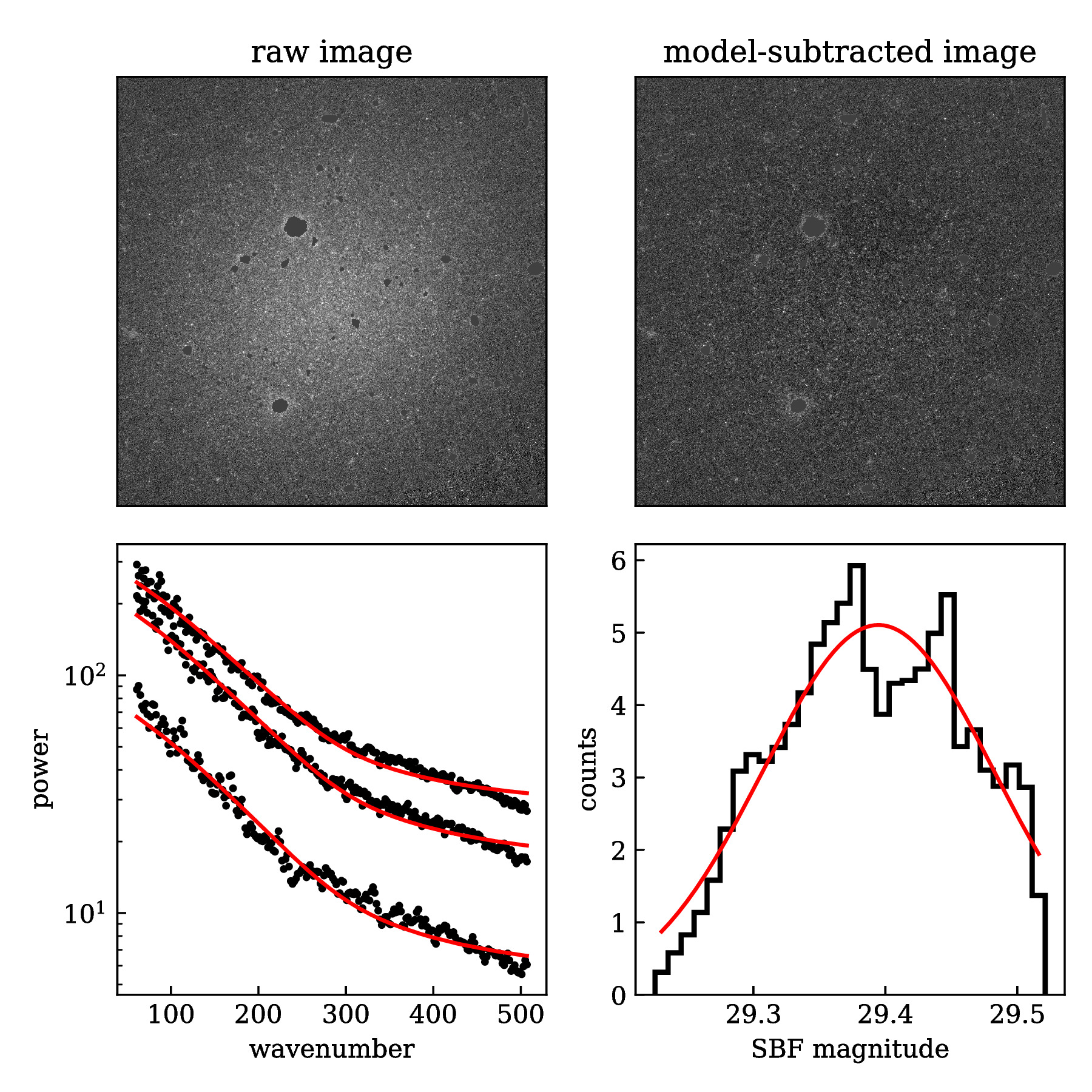}
\caption{Fits to the azimuthally averaged power spectra for NGC\,1052-DF2 in three annuli, as described in the text, showing the section of the original (masked) image within which the measurement is performed (top left), the same section of the image after subtracting the GALFIT model (top right), power spectrum measurements of the data within each annulus and best fit curves (bottom left), and the distribution of SBF magnitudes from numerous trials of varying the fitting range (bottom right). The adopted SBF magnitude and uncertainty are the mean and spread of the Gaussian fit to the histogram. \label{fig:sbf}}
\end{figure}

The other distance estimation method we use is the surface brightness fluctuations (SBF) method.
The goal is to measure distances to galaxies that are beyond the reach of the TRGB method (with our photometric depth), but we apply it to as many galaxies in the sample as possible so that we can test whether the two methods are consistent with each other.
The SBF distance indicator was originally described by \cite{tonry} and has since been used widely.
In this method, one measures the pixel-to-pixel variance in the surface brightness a galaxy arising from fluctuations in the number of stars per resolution element; the amplitude of the measured fluctuations scales inversely with the distance to the galaxy \citep{BM}.
The metric is the SBF magnitude, which is effectively the luminosity-weighted average brightness of the stars in the galaxy.

We followed most of the usual prescription for applying this method \citep[e.g.][]{mei2005}.
First, we apply the masks used in the GALFIT fitting to the drizzled $I_{814}$ images.
Then, we subtract a smooth model of the galaxy.
Other authors have used elliptical isophotes for the model but here we simply use the best-fitting GALFIT models.
Next, we divide the image into concentric elliptical annuli, with position angle and axis ratio corresponding to the GALFIT model, and measure the 2-D power spectrum within each annulus.
We do not attempt to separately model and remove contributions to the SBF from globular clusters and background objects \citep[see][]{mei2005} because they were already included in the mask and/or their contribution to the uncertainty in the measurement is small compared to the random uncertainties from the fits.
Next, we measure the power spectrum of the normalized point spread function (PSF) and of the window function, where the window function is the product of the mask, the square root of the smooth model, and an annular mask which selects the region being analyzed, in each concentric annulus.
We construct the PSF from unsaturated stars in the field.
For each annulus, the image power spectrum $P(k)$ is then modeled by a linear function:
$$ P(k) = P_0 E(k) + P_1 $$
where $E(k)$ is the ``expectation power spectrum'', which is the convolution of the normalized PSF power spectrum and the window function power spectrum, and $P_0$ and $P_1$ are constants which are optimized to provide the best fit.
The constant $P_1$ represents the flat, white-noise component of the power spectrum.
The constant $P_0$ is the quantity of interest; it is the amplitude of surface brightness fluctuations in units of the original image, which we then convert into the corresponding magnitude.

In practice, one usually fits the azimuthally averaged power spectra, as shown in Figure \ref{fig:sbf}.
For each annulus, we performed the fit multiple times, allowing the range of $k$ values over which the fit is performed to vary slightly until convergence was reached.
The reason for allowing this flexibility is because at low wave numbers, the power spectra may be affected by large-scale background light subtraction, while at high wave numbers, the spectra may be affected by pixel correlations introduced by the geometric correction from drizzling the images.
The section of the image used in the measurement was manually tweaked for each galaxy in order to increase the stability of the fits; it was typically a region of radius between $0.5 R_e$ to $1.5 R_e$ centered on the galaxy.
The annuli were then excised from within this region.

\begin{figure*}[]
\plotone{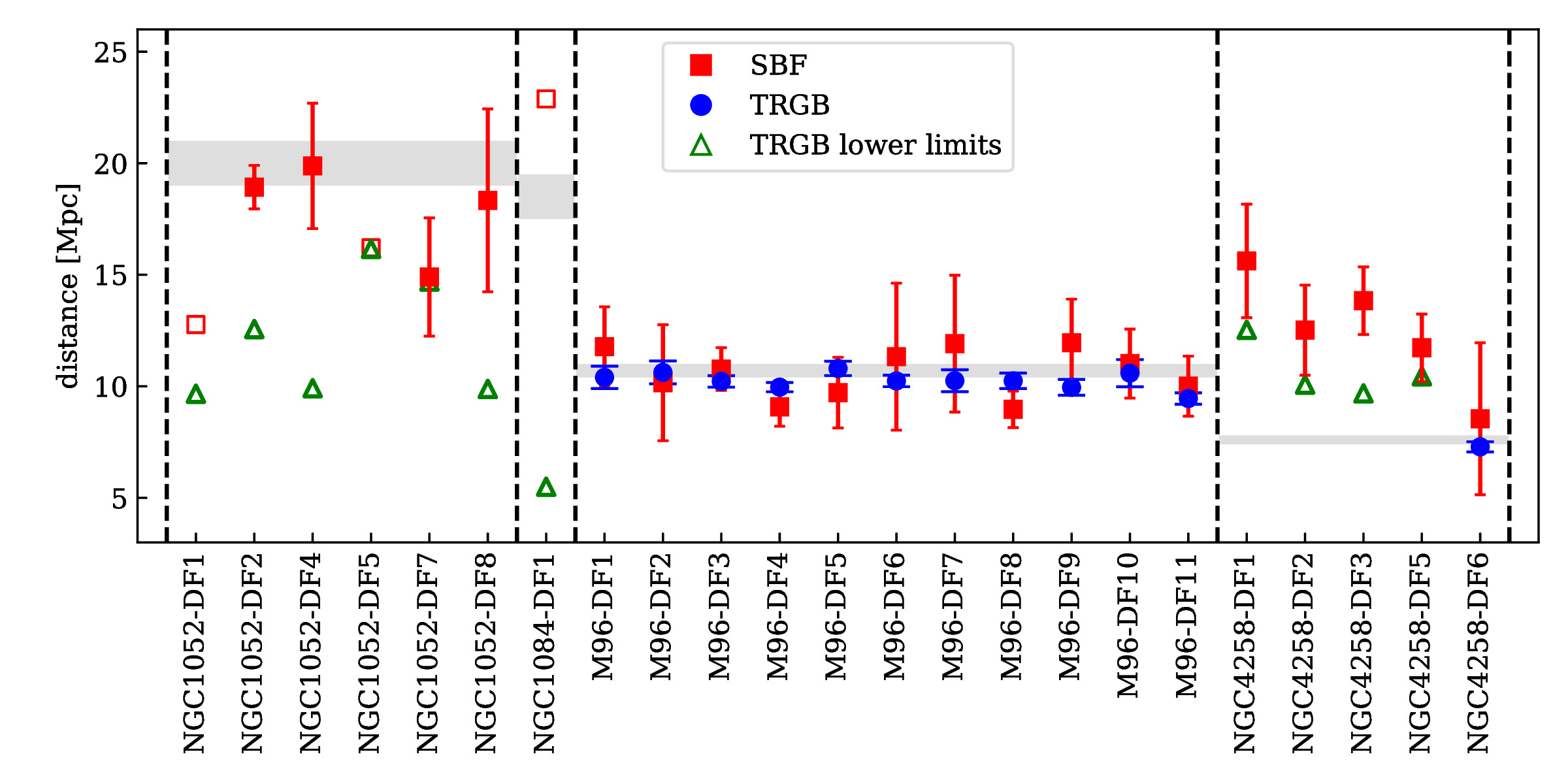}
\caption{
Distance estimates to the galaxies in the sample, derived from two different methods, as described in the text.
The red squares are from the surface brightness fluctuation (SBF) method, the blue circles from the tip of the red giant branch (TRGB) method, and the green triangles indicate approximate lower limits derived from non-detections of the tip of the red giant branch (TRGB).
The open squares correspond to objects with suspect SBF measurements due to extremely low surface brightness or large contamination from blue stars.
Note that the error bars on both the SBF and TRGB distances represent the random errors associated with the measurement and do not take into account systematic errors owing to calibration uncertainties, etc.
The different fields are delimited by the black dashed lines.
The gray shaded regions represent the distance to the main galaxy/group in the field, as quoted in Section \ref{sec:HST imaging}.
The SBF and TRGB methods are consistent with each other, and the distances of the majority of the galaxies are consistent with the grey bands.
Note that several of the galaxies in the NGC\,4258 field do not belong to the NGC\,4258 group, but are located in the background.
\label{fig:distcompare}}
\end{figure*}

The uncertainties are dominated by the variance in the measured SBF magnitude in the different annuli.
In order to estimate the random errors from the fits, we recorded the SBF magnitude measured in each annulus and for each range of wave numbers fit.
We then generated a histogram of the resulting magnitudes, weighted by the $\chi^2$ values of the fits, and fit the peak with a Gaussian.
We take the mean and spread of this Gaussian to be the SBF magnitude and its uncertainty.
In all cases, we verified the results by examining diagnostic plots like the ones shown in Figure \ref{fig:sbf}.
We find that the convergence and stability of the SBF measurement depends on the surface brightness of the galaxy.
For this reason, we caution that the measurements for the most diffuse objects ($\mu_{0,V} > 25.5$) in the sample may be suspect.
The other complication pertains to the galaxy NGC\,1084-DF1, which has significant clumpy structure, so the smooth GALFIT model is not adequate for producing a clean model-subtracted image.
The clumps correspond to star forming regions, which presents another complication as main sequence stars are contaminating the SBF signal. 
In the interest of consistency we still perform the SBF measurement (with smooth model and extrapolation of the calibration) on this galaxy, but caution that the result is suspect.
We indicate the corresponding distance estimates from these suspect measurements as open squares in Figure \ref{fig:distcompare}.

Distance estimates were calculated by drawing 1000 magnitude values from the Gaussian used to fit the SBF magnitude, and assuming a color-dependent absolute magnitude from the calibrated relation of \cite{blakeslee}, reproduced here:
\begin{equation*}
\begin{split}
\overline{M}_{814} = (-1.168 \pm 0.013 \pm 0.092) \\
+ (1.83 \pm 0.2) [(g_{475} - I_{814}) - 1.2].
\end{split}
\end{equation*}
To obtain the required color index, we derive the relation $g_{475} - I_{814} = 1.852(V_{606} - I_{814}) + 0.096$ from synthetic photometry and use the overall $V_{606}-I_{814}$ color of the galaxy, reported in Table \ref{tab:gftable}\footnote{We note that the $g_{475} - I_{814}$ color of most of the galaxies in our sample lie outside of the range that the \cite{blakeslee} relation is calibrated for, but we find that an extrapolation may still be valid \citep[see Fig. 4 of][]{vd:2018d}}.
We take the mean and standard deviation of the resulting distribution of distance estimates as the distance estimate and random uncertainty, plotted as red squares in Figure \ref{fig:distcompare}.

As noted above, we performed the SBF measurement on all the galaxies, even those that have a distance measurement from the location of the TRGB.
The SBF measurements are challenging for galaxies of such low surface brightness, and there is certainly room for improvement in our methodology.
Despite that, there is generally good agreement between the SBF and TRGB methods, and the differences are generally consistent with the estimated uncertainties, as seen in Figure \ref{fig:distcompare}.
Furthermore, the SBF distances estimated for the objects with no detectable TRGB are consistent with the lower limits from the TRGB non-detection.

\begin{deluxetable}{lll}[]
\tabletypesize{\scriptsize}
\tablecaption{Distances shown in Figure \ref{fig:distcompare} \label{tab:distances}}
\tablenum{2}
\tablehead{\colhead{Id} & \colhead{TRGB distance} & \colhead{SBF distance} \vspace{-0.2cm} \\
\colhead{} & \colhead{(Mpc)} & \colhead{(Mpc)} 
}
\startdata
NGC1052-DF1 & $> 9.9$   & 12.4: \\
NGC1052-DF2 & $> 12.6$   & 18.9 $\pm$ 1.0 \\
NGC1052-DF4 & $> 9.7$    & 19.9 $\pm$ 2.8 \\
NGC1052-DF5 & $> 16.2$   & 16.2: \\
NGC1052-DF7 & $> 14.7$   & 14.9 $\pm$ 2.6 \\
NGC1052-DF8 & $> 9.8$   & 18.3 $\pm$ 4.1 \\
NGC1084-DF1 & $> 5.5$    & 22.9: \\
M96-DF1 & 10.4 $\pm$ 0.5 & 11.8 $\pm$ 1.8 \\
M96-DF2 & 10.6 $\pm$ 0.5 & 10.2 $\pm$ 2.6 \\
M96-DF3 & 10.2 $\pm$ 0.3 & 10.8 $\pm$ 1.0 \\
M96-DF4 & 10.0 $\pm$ 0.2 & 9.1 $\pm$ 0.9 \\
M96-DF5 & 10.8 $\pm$ 0.3 & 9.7 $\pm$ 1.6 \\
M96-DF6 & 10.2 $\pm$ 0.3 & 11.3 $\pm$ 3.3 \\
M96-DF7 & 10.2 $\pm$ 0.5 & 11.9 $\pm$ 3.1 \\
M96-DF8 & 10.2 $\pm$ 0.3 & 9.0 $\pm$ 0.8 \\
M96-DF9 & 10.0 $\pm$ 0.4 & 12.0 $\pm$ 2.0 \\
M96-DF10 & 10.6 $\pm$ 0.6 & 11.0 $\pm$ 1.5 \\
M96-DF11 & 9.5 $\pm$ 0.3 & 10.0 $\pm$ 1.3 \\
NGC4258-DF1 & $> 12.5$   & 15.6 $\pm$ 2.5 \\
NGC4258-DF2 & $> 10.1$    & 12.5 $\pm$ 2.0 \\
NGC4258-DF3 & $> 9.7$   & 13.8 $\pm$ 1.5 \\
NGC4258-DF5 & $> 10.4$   & 11.7 $\pm$ 1.5 \\
NGC4258-DF6 & 7.3 $\pm$ 0.2 & 8.5 $\pm$ 3.4 \\
\enddata
\tablecomments{Lower distance limits from the TRGB non-detection are given where applicable. See Section \ref{sec:trgb} for an explanation.
SBF distances marked with a colon are highly uncertain due to extremely low central surface brightness or too much contamination from blue stars.}
\end{deluxetable}

\begin{figure*}[]
\epsscale{1.}
\plotone{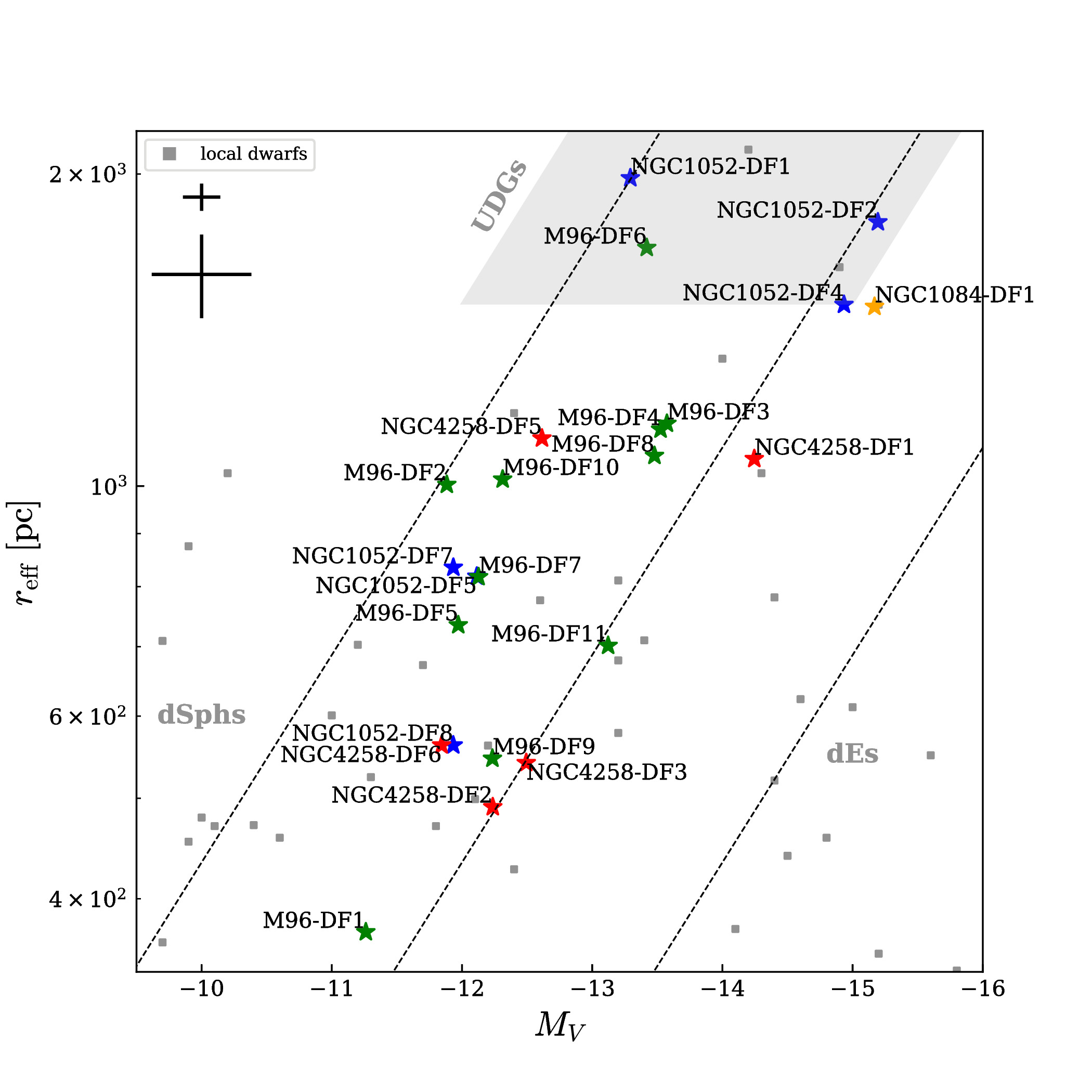}
\caption{Size--luminosity relation for our sample (colored stars), dwarf galaxies in and around the Local Group (gray squares; \cite{mccon2012}), and ultra--diffuse galaxies (UDGs; grey shaded region; \cite{vd2015a}).
The colors of the stars are redundant with the name of the field that the galaxies belong to.
Wherever available, we use the TRGB distance to derive the physical size and luminosity, and the SBF distance otherwise.
The black crosses under the legend represent the typical uncertainties on the physical size and absolute magnitude from TRGB distances (top) and SBF distances (bottom).
The dashed lines are lines of constant effective surface brightness at $\mu_{e,V}=$ 22, 24, and 26.
\label{fig:sizelum}
}
\end{figure*}

\section{Discussion} \label{sec:discussion}

\subsection{Environment}

As discussed in Section \ref{sec:intro}, our analysis of low surface brightness galaxies in the M101 field showed that only three of the seven are members of the M101 group (at $\sim 7$\,Mpc) \citep{danieli2016} with the other four likely associated with the background NGC\,5485 group at $\sim 27$\,Mpc \citep{merritt2016}.
As for the fields examined in this work, we can glean the following from Figure \ref{fig:distcompare}:
All 11 galaxies in the M\,96 field are at the distance of the M\,96 group.
At least three of the galaxies in the NGC\,1052 field are consistent with the distance to NGC\,1052 within errors, and none have a resolved RGB in their CMDs from two-orbit ACS imaging, placing them all at $\gtrsim 14$\,Mpc (which means that they could all be associated with the NGC\,1052 group).
In the NGC\,4258 field, only one galaxy is at the distance of NGC\,4258 with the other four in the background at $\gtrsim 12$\,Mpc.
It is not clear which, if any, background galaxies or group(s) these galaxies could be associated with; this is a rich area of the sky with many fairly bright galaxies at different distances. 
It remains an open question at this time whether they are all satellites of background galaxies or one or more is an isolated (central) low mass quiescent galaxy.
Deeper imaging for accurate TRGB distances, as well as radial velocity measurements, could shed more light on this important question.
Lastly, it is difficult to comment on the group membership of NGC\,1084-DF1 due to its highly uncertain distance.
Overall, we infer from our small sample that $\sim 75$\,\% of faint low surface brightness ``blobs'' (15/20 with secure distances) are likely associated with the targeted groups, for the particular combination of apparent sizes, surface brightnesses, and group distances that are probed in this study.

\subsection{Sizes}
The derived distances allow the effective radii of the galaxies to be expressed in physical units. 
In Figure \ref{fig:sizelum} we show our sample of galaxies in the size--luminosity plane, along with dwarf galaxies in and around the Local Group \citep{mccon2012}.
Wherever available, we use the TRGB distance to derive the physical size and luminosity, and the SBF distance otherwise. 
For this Figure we assume that the three galaxies without stable distance measurements are at the distance of the targeted group.
The red shaded region shows where UDGs \citep{vd2015a} fall within this parameter space.
Four of the galaxies are (nominally) UDGs: the well-studied galaxy NGC\,1052-DF2 \citep{df2nature,df2apj,df2rn}; NGC\,1052-DF4, which is a similar-looking galaxy in the same group, the extremely LSB galaxy NGC\,1052-DF1, which has a suspect distance measurement (and thus also physical size),  and the somewhat irregular galaxy M\,96-DF6.
The latter galaxy is one of the closest UDGs currently known: with a TRGB distance of $10.2 \pm 3$\,Mpc, it may be slightly closer than UGC\,2162 \citep{trujillo2017}\footnote{The galaxy F8D1 in the M81 group is even closer \citep{caldwell98}. Furthermore, the Andromeda satellite And\,XIX could also be considered a UDG \citep{mccon08}.}.

Most of the other objects fall somewhere in between Local Group dwarf spheroidals (dSphs) and UDGs in terms of their size and luminosity.
There are only a few galaxies in the Local Group with $R_e \sim 1$\,kpc and $M_V \sim -12$.
By observing more groups, and \textit{more massive} groups such as M\,96, we are able to fill in this region of parameter space.

\begin{figure}[]
\epsscale{1.1}
\plotone{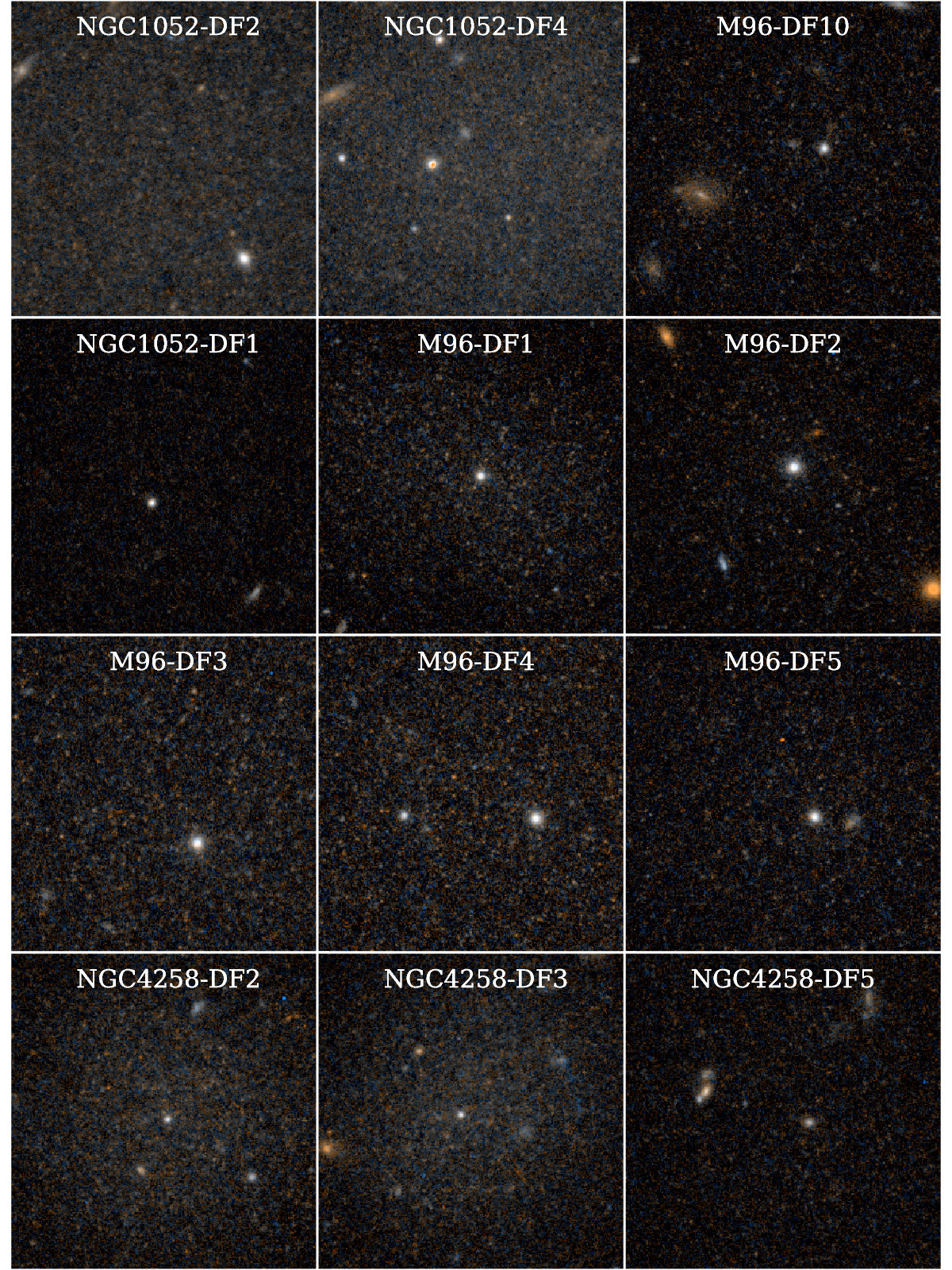}
\caption{Pseudo-color \textit{HST}/ACS images of the central $15''$ of twelve galaxies in the sample which appear to contain a bright, compact object(s) within this region. \label{fig:nucleatedblobs}
}
\end{figure}

\subsection{Nuclear star clusters and globular clusters \label{sec:starclusters}}
Several of the galaxies feature a nuclear star cluster or globular cluster system.
The unusual globular cluster system of NGC\,1052-DF2 has been studied in detail by \cite{df2apj}.
NGC\,1052-DF4 also appears to host a globular cluster population.
Shown in Figure \ref{fig:nucleatedblobs} are color images of the central $15''$ of twelve of the galaxies in the sample that seem to have (at least one) bright compact object within this region.
These compact objects all have color $V_{\rm 606} - I_{\rm 814} \sim 0.37$ and absolute magnitude $M_{I_{\rm 814}} \sim -8.5$, which is nearly 1 magnitude brighter than the peak of the globular cluster luminosity function.
A detailed analysis of their properties is beyond the scope of this paper; here we note that these compact star clusters appear to be quite common in large, low surface brightness galaxies \citep[][and references therein]{Koda:2015,lim2018}.

\section{Conclusions and Future Work}

In this work, we present a sample of  low surface brightness galaxies ($\mu_{e,V} \sim 25 - 27 \, \mathrm{mag \, arcsec^{-2}})$ identified with the Dragonfly Telephoto Array and subsequently followed-up with \textit{HST}/ACS $V_{606}$ and $I_{814}$ imaging.
Several of the galaxies are new discoveries, while others had been detected as candidate dwarfs by previous surveys, though none have had an accurate distance measurement.
We measure their distances using the TRGB method where applicable and the SBF method otherwise.
The SBF measurement is difficult for such diffuse galaxies, but we show that the estimates obtained using this method agree within the uncertainties with those obtained using the TRGB method, in the distance regime where both are applicable.

From the distance estimates, we derive physical sizes and luminosities and find that the galaxies are mostly large ($R \sim 1\, \mathrm{kpc}$), luminous ($M_V \sim -12.5$) dwarfs.
A few are large and diffuse enough to be considered UDGs, which are then among the nearest known UDGs at the time of this writing.
Most, if not all, of the galaxies in the NGC\,1052, NGC\,1084, and M\,96 fields are consistent with being at the distance of those groups.
The discovery of several new dwarfs in the M\,96 group means that the LF of this group can now be updated to include these objects and the question of the `missing satellites' and `too big to fail' problems in this group can be reevaluated.
On the other hand, four out of five of the galaxies in the field of NGC\,4258 ($\sim 7\, \mathrm{Mpc}$) appear to be in the background at roughly twice the distance of the NGC\,4258 group.
Radial velocity measurements will be helpful in order to determine which, if any, galaxies these objects are satellites of.

\textit{HST} imaging reveals that the low surface brightness galaxies exhibit a number of interesting features.
About half of them appear to host a globular cluster system and/or a large, nuclear star cluster.
Among them is NGC\,1052-DF2, which has recently been shown to be highly dark matter deficient from kinematic observations of its globular clusters \citep{df2nature}; the clusters themselves also appear highly unusual in their brightness and structure \citep{df2apj}.
Another galaxy in this group, NGC\,1052-DF4, appears to host a similar, albeit slightly less numerous, population of clusters and we plan to obtain kinematic measurements of them in order to constrain the galaxy's mass.
Two of the galaxies, NGC\,1052-DF8 and NGC\,1084-DF1, appear to have active star-forming regions, while the rest have old, red stellar populations.

This paper presents a first exploration, and we are planning further studies of these objects.
We conclude by highlighting the synergy between low surface brightness imaging surveys and HST.
Telescopes such as Dragonfly have the capacity to quickly and efficiently detect extended low surface brightness objects, and HST follow-up can then be used to study them in exquisite detail.
We plan to search for more of these objects as part of the ongoing Dragonfly Wide Field Survey \citep{danieli18}.

\section*{Acknowledgements}
We thank Marijn Franx for comments on the manuscript.
YC thanks the Gruber Foundation and Patricia Gruber for the Gruber Fellowship, which supported this work.
Support from STScI grants HST-GO-13682 and HST-GO-14644, as well as NSF grants AST-1312376 and AST-1613582, is gratefully acknowledged.
J.M.D.K. gratefully acknowledges funding from the German Research Foundation (DFG) in the form of an Emmy Noether Research Group (grant number KR4801/1-1) and from the European Research Council (ERC) under the European Union's Horizon 2020 research and innovation programme via the ERC Starting Grant MUSTANG (grant agreement number 714907).
AJR was supported by NSF grants AST-1515084 and AST-1616710, and as a Research Corporation for Science Advancement Cottrell Scholar.
This research has made use of NASA's Astrophysics Data System.
We acknowledge the use of the open source Python packages Astropy, Numpy, Scipy, Matplotlib, Pandas, and IPython.

\clearpage

\bibliography{ref}

\begin{thebibliography}{}
\expandafter\ifx\csname natexlab\endcsname\relax\def\natexlab#1{#1}\fi

\bibitem[{{Abraham} {et~al.}(2018){Abraham}, {Danieli}, {van Dokkum}, {Conroy},
  {Kruijssen}, {Cohen}, {Merritt}, {Zhang}, {Lokhorst}, {Mowla}, {Brodie},
  {Romanowsky}, \& {Janssens}}]{df2stream}
{Abraham}, R., {Danieli}, S., {van Dokkum}, P., {et~al.} 2018, Research Notes
  of the American Astronomical Society, 2, 16

\bibitem[{{Abraham} \& {van Dokkum}(2014)}]{dragonfly}
{Abraham}, R.~G., \& {van Dokkum}, P.~G. 2014, Publications of the Astronomical
  Society of the Pacific, 126, 55

\bibitem[{{Barbarino} {et~al.}(2015){Barbarino}, {Dall'Ora}, {Botticella},
  {Della Valle}, {Zampieri}, {Maund}, {Pumo}, {Jerkstrand}, {Benetti},
  {Elias-Rosa}, {Fraser}, {Gal-Yam}, {Hamuy}, {Inserra}, {Knapic}, {LaCluyze},
  {Molinaro}, {Ochner}, {Pastorello}, {Pignata}, {Reichart}, {Ries},
  {Riffeser}, {Schmidt}, {Schmidt}, {Smareglia}, {Smartt}, {Smith},
  {Sollerman}, {Sullivan}, {Tomasella}, {Turatto}, {Valenti}, {Yaron}, \&
  {Young}}]{Barbarino:2015}
{Barbarino}, C., {Dall'Ora}, M., {Botticella}, M.~T., {et~al.} 2015, \mnras,
  448, 2312

\bibitem[{{Beasley} \& {Trujillo}(2016)}]{BT16}
{Beasley}, M.~A., \& {Trujillo}, I. 2016, \apj, 830, 23

\bibitem[{{Belokurov} {et~al.}(2007){Belokurov}, {Zucker}, {Evans}, {Kleyna},
  {Koposov}, {Hodgkin}, {Irwin}, {Gilmore}, {Wilkinson}, {Fellhauer},
  {Bramich}, {Hewett}, {Vidrih}, {De Jong}, {Smith}, {Rix}, {Bell}, {Wyse},
  {Newberg}, {Mayeur}, {Yanny}, {Rockosi}, {Gnedin}, {Schneider}, {Beers},
  {Barentine}, {Brewington}, {Brinkmann}, {Harvanek}, {Kleinman}, {Krzesinski},
  {Long}, {Nitta}, \& {Snedden}}]{Belokurov:2007}
{Belokurov}, V., {Zucker}, D.~B., {Evans}, N.~W., {et~al.} 2007, \apj, 654, 897

\bibitem[{{Bertin} \& {Arnouts}(1996)}]{sextractor}
{Bertin}, E., \& {Arnouts}, S. 1996, Astronomy and Astrophysics Supplement
  Series, 117, 393

\bibitem[{{Binggeli} {et~al.}(1990){Binggeli}, {Tarenghi}, \& {Sandage}}]{BTS}
{Binggeli}, B., {Tarenghi}, M., \& {Sandage}, A. 1990, \aap, 228, 42

\bibitem[{{Binney} \& {Merrifield}(1998)}]{BM}
{Binney}, J., \& {Merrifield}, M. 1998, {Galactic Astronomy}

\bibitem[{{Blakeslee} {et~al.}(2001){Blakeslee}, {Lucey}, {Barris}, {Hudson},
  \& {Tonry}}]{Blakeslee:2001}
{Blakeslee}, J.~P., {Lucey}, J.~R., {Barris}, B.~J., {Hudson}, M.~J., \&
  {Tonry}, J.~L. 2001, \mnras, 327, 1004

\bibitem[{{Blakeslee} {et~al.}(2010){Blakeslee}, {Cantiello}, {Mei},
  {C{\^o}t{\'e}}, {Barber DeGraaff}, {Ferrarese}, {Jord{\'a}n}, {Peng},
  {Tonry}, \& {Worthey}}]{blakeslee}
{Blakeslee}, J.~P., {Cantiello}, M., {Mei}, S., {et~al.} 2010, \apj, 724, 657

\bibitem[{{Bothun} {et~al.}(1987){Bothun}, {Impey}, {Malin}, \&
  {Mould}}]{bothun87}
{Bothun}, G.~D., {Impey}, C.~D., {Malin}, D.~F., \& {Mould}, J.~R. 1987, \aj,
  94, 23

\bibitem[{{Caldwell} {et~al.}(1998){Caldwell}, {Armandroff}, {Da Costa}, \&
  {Seitzer}}]{caldwell98}
{Caldwell}, N., {Armandroff}, T.~E., {Da Costa}, G.~S., \& {Seitzer}, P. 1998,
  \aj, 115, 535

\bibitem[{{Dalcanton} {et~al.}(2009){Dalcanton}, {Williams}, {Seth}, {Dolphin},
  {Holtzman}, {Rosema}, {Skillman}, {Cole}, {Girardi}, {Gogarten},
  {Karachentsev}, {Olsen}, {Weisz}, {Christensen}, {Freeman}, {Gilbert},
  {Gallart}, {Harris}, {Hodge}, {de Jong}, {Karachentseva}, {Mateo}, {Stetson},
  {Tavarez}, {Zaritsky}, {Governato}, \& {Quinn}}]{ANGST}
{Dalcanton}, J.~J., {Williams}, B.~F., {Seth}, A.~C., {et~al.} 2009, The
  Astrophysical Journal Supplement Series, 183, 67

\bibitem[{{Danieli} {et~al.}(2018){Danieli}, {van Dokkum}, \&
  {Conroy}}]{danieli18}
{Danieli}, S., {van Dokkum}, P., \& {Conroy}, C. 2018, \apj, 856, 69

\bibitem[{{Danieli} {et~al.}(2017){Danieli}, {van Dokkum}, {Merritt},
  {Abraham}, {Zhang}, {Karachentsev}, \& {Makarova}}]{danieli2016}
{Danieli}, S., {van Dokkum}, P., {Merritt}, A., {et~al.} 2017, \apj, 837, 136

\bibitem[{{Di Cintio} {et~al.}(2017){Di Cintio}, {Brook}, {Dutton},
  {Macci{\`o}}, {Obreja}, \& {Dekel}}]{dicintio17}
{Di Cintio}, A., {Brook}, C.~B., {Dutton}, A.~A., {et~al.} 2017, \mnras, 466,
  L1

\bibitem[{{Dolphin}(2000)}]{dolphot}
{Dolphin}, A.~E. 2000, Publications of the Astronomical Society of the Pacific,
  112, 1383

\bibitem[{{Graham} \& {Driver}(2005)}]{sersic}
{Graham}, A.~W., \& {Driver}, S.~P. 2005, Publications of the Astronomical
  Society of Australia, 22, 118

\bibitem[{{Humphreys} {et~al.}(2013){Humphreys}, {Reid}, {Moran}, {Greenhill},
  \& {Argon}}]{Humphreys:2013}
{Humphreys}, E.~M.~L., {Reid}, M.~J., {Moran}, J.~M., {Greenhill}, L.~J., \&
  {Argon}, A.~L. 2013, \apj, 775, 13

\bibitem[{{Impey} {et~al.}(1996){Impey}, {Sprayberry}, {Irwin}, \&
  {Bothun}}]{impey96}
{Impey}, C.~D., {Sprayberry}, D., {Irwin}, M.~J., \& {Bothun}, G.~D. 1996, The
  Astrophysical Journal Supplement Series, 105, 209

\bibitem[{{Karachentsev} {et~al.}(2014){Karachentsev}, {Kaisina}, \&
  {Makarov}}]{kkm2014}
{Karachentsev}, I.~D., {Kaisina}, E.~I., \& {Makarov}, D.~I. 2014, \aj, 147, 13

\bibitem[{{Karachentsev} \& {Karachentseva}(2004)}]{kk2004}
{Karachentsev}, I.~D., \& {Karachentseva}, V.~E. 2004, Astronomy Reports, 48,
  267

\bibitem[{{Karachentsev} {et~al.}(2000){Karachentsev}, {Karachentseva},
  {Suchkov}, \& {Grebel}}]{kks2000}
{Karachentsev}, I.~D., {Karachentseva}, V.~E., {Suchkov}, A.~A., \& {Grebel},
  E.~K. 2000, Astronomy and Astrophysics Supplement Series, 145, 415

\bibitem[{{Karachentsev} {et~al.}(2013){Karachentsev}, {Makarov}, \&
  {Kaisina}}]{kmk2013}
{Karachentsev}, I.~D., {Makarov}, D.~I., \& {Kaisina}, E.~I. 2013, \aj, 145,
  101

\bibitem[{{Karachentsev} {et~al.}(2017){Karachentsev}, {Makarova}, {Sharina},
  \& {Karachentseva}}]{kara2017}
{Karachentsev}, I.~D., {Makarova}, L.~N., {Sharina}, M.~E., \& {Karachentseva},
  V.~E. 2017, Astrophysical Bulletin, 72, 376

\bibitem[{{Karachentsev} {et~al.}(2015){Karachentsev}, {Riepe}, {Zilch},
  {Blauensteiner}, {Elvov}, {Hochleitner}, {Hubl}, {Kerschhuber},
  {K{\"u}ppers}, {Neyer}, {P{\"o}lzl}, {Remmel}, {Schneider}, {Sparenberg},
  {Trulson}, {Willems}, \& {Ziegler}}]{krz2015}
{Karachentsev}, I.~D., {Riepe}, P., {Zilch}, T., {et~al.} 2015, Astrophysical
  Bulletin, 70, 379

\bibitem[{{Karachentseva} \& {Karachentsev}(1998)}]{kk1998}
{Karachentseva}, V.~E., \& {Karachentsev}, I.~D. 1998, Astronomy and
  Astrophysics Supplement Series, 127, 409

\bibitem[{{Kim} {et~al.}(2011){Kim}, {Kim}, {Hwang}, {Lee}, {Chun}, \&
  {Ann}}]{kkh2011}
{Kim}, E., {Kim}, M., {Hwang}, N., {et~al.} 2011, \mnras, 412, 1881

\bibitem[{{Koda} {et~al.}(2015){Koda}, {Yagi}, {Yamanoi}, \&
  {Komiyama}}]{Koda:2015}
{Koda}, J., {Yagi}, M., {Yamanoi}, H., \& {Komiyama}, Y. 2015, \apj, 807, L2

\bibitem[{{Laporte} {et~al.}(2018){Laporte}, {Agnello}, \&
  {Navarro}}]{laporte18}
{Laporte}, C. F.~P., {Agnello}, A., \& {Navarro}, J.~F. 2018, ArXiv e-prints,
  arXiv:1804.04139

\bibitem[{{Lim} {et~al.}(2018){Lim}, {Peng}, {Cote}, {Sales}, {den Brok},
  {Blakeslee}, \& {Guhathakurta}}]{lim2018}
{Lim}, S., {Peng}, E.~W., {Cote}, P., {et~al.} 2018, ArXiv e-prints,
  arXiv:1806.05425

\bibitem[{{Lupton} {et~al.}(2004){Lupton}, {Blanton}, {Fekete}, {Hogg},
  {O'Mullane}, {Szalay}, \& {Wherry}}]{lupton}
{Lupton}, R., {Blanton}, M.~R., {Fekete}, G., {et~al.} 2004, Publications of
  the Astronomical Society of the Pacific, 116, 133

\bibitem[{{Makarov} {et~al.}(2006){Makarov}, {Makarova}, {Rizzi}, {Tully},
  {Dolphin}, {Sakai}, \& {Shaya}}]{makarov}
{Makarov}, D., {Makarova}, L., {Rizzi}, L., {et~al.} 2006, \aj, 132, 2729

\bibitem[{{Martin} {et~al.}(2018){Martin}, {Collins}, {Longeard}, \&
  {Tollerud}}]{martin18}
{Martin}, N.~F., {Collins}, M. L.~M., {Longeard}, N., \& {Tollerud}, E. 2018,
  \apj, 859, L5

\bibitem[{{Martin} {et~al.}(2013){Martin}, {Slater}, {Schlafly}, {Morganson},
  {Rix}, {Bell}, {Laevens}, {Bernard}, {Ferguson}, {Finkbeiner}, {Burgett},
  {Chambers}, {Hodapp}, {Kaiser}, {Kudritzki}, {Magnier}, {Morgan}, {Price},
  {Tonry}, \& {Wainscoat}}]{Martin:2013}
{Martin}, N.~F., {Slater}, C.~T., {Schlafly}, E.~F., {et~al.} 2013, \apj, 772,
  15

\bibitem[{{McConnachie}(2012)}]{mccon2012}
{McConnachie}, A.~W. 2012, \aj, 144, 4

\bibitem[{{McConnachie} {et~al.}(2008){McConnachie}, {Huxor}, {Martin},
  {Irwin}, {Chapman}, {Fahlman}, {Ferguson}, {Ibata}, {Lewis}, {Richer}, \&
  {Tanvir}}]{mccon08}
{McConnachie}, A.~W., {Huxor}, A., {Martin}, N.~F., {et~al.} 2008, \apj, 688,
  1009

\bibitem[{{McQuinn} {et~al.}(2017){McQuinn}, {Skillman}, {Dolphin}, {Berg}, \&
  {Kennicutt}}]{mcquinn}
{McQuinn}, K. B.~W., {Skillman}, E.~D., {Dolphin}, A.~E., {Berg}, D., \&
  {Kennicutt}, R. 2017, \aj, 154, 51

\bibitem[{{Mei} {et~al.}(2005){Mei}, {Blakeslee}, {Tonry}, {Jord{\'a}n},
  {Peng}, {C{\^o}t{\'e}}, {Ferrarese}, {West}, {Merritt}, \&
  {Milosavljevi{\'c}}}]{mei2005}
{Mei}, S., {Blakeslee}, J.~P., {Tonry}, J.~L., {et~al.} 2005, \apj, 625, 121

\bibitem[{{M{\'e}ndez} {et~al.}(2002){M{\'e}ndez}, {Davis}, {Moustakas},
  {Newman}, {Madore}, \& {Freedman}}]{mendez}
{M{\'e}ndez}, B., {Davis}, M., {Moustakas}, J., {et~al.} 2002, \aj, 124, 213

\bibitem[{{Merritt} {et~al.}(2014){Merritt}, {van Dokkum}, \&
  {Abraham}}]{merritt2014}
{Merritt}, A., {van Dokkum}, P., \& {Abraham}, R. 2014, \apj, 787, L37

\bibitem[{{Merritt} {et~al.}(2016{\natexlab{a}}){Merritt}, {van Dokkum},
  {Abraham}, \& {Zhang}}]{Merritt:2016}
{Merritt}, A., {van Dokkum}, P., {Abraham}, R., \& {Zhang}, J.
  2016{\natexlab{a}}, \apj, 830, 62

\bibitem[{{Merritt} {et~al.}(2016{\natexlab{b}}){Merritt}, {van Dokkum},
  {Danieli}, {Abraham}, {Zhang}, {Karachentsev}, \& {Makarova}}]{merritt2016}
{Merritt}, A., {van Dokkum}, P., {Danieli}, S., {et~al.} 2016{\natexlab{b}},
  \apj, 833, 168

\bibitem[{{Mihos} {et~al.}(2015){Mihos}, {Durrell}, {Ferrarese}, {Feldmeier},
  {C{\^o}t{\'e}}, {Peng}, {Harding}, {Liu}, {Gwyn}, \& {Cuillandre}}]{mihos15}
{Mihos}, J.~C., {Durrell}, P.~R., {Ferrarese}, L., {et~al.} 2015, \apj, 809,
  L21

\bibitem[{{Mowla} {et~al.}(2017){Mowla}, {van Dokkum}, {Merritt}, {Abraham},
  {Yagi}, \& {Koda}}]{mowla17}
{Mowla}, L., {van Dokkum}, P., {Merritt}, A., {et~al.} 2017, \apj, 851, 27

\bibitem[{{Mu{\~n}oz} {et~al.}(2015){Mu{\~n}oz}, {Eigenthaler}, {Puzia},
  {Taylor}, {Ordenes-Brice{\~n}o}, {Alamo-Mart{\'\i}nez}, {Ribbeck},
  {{\'A}ngel}, {Capaccioli}, {C{\^o}t{\'e}}, {Ferrarese}, {Galaz}, {Hempel},
  {Hilker}, {Jord{\'a}n}, {Lan{\c{c}}on}, {Mieske}, {Paolillo}, {Richtler},
  {S{\'a}nchez-Janssen}, \& {Zhang}}]{munoz15}
{Mu{\~n}oz}, R.~P., {Eigenthaler}, P., {Puzia}, T.~H., {et~al.} 2015, \apj,
  813, L15

\bibitem[{{M{\"u}ller} {et~al.}(2018){M{\"u}ller}, {Jerjen}, \&
  {Binggeli}}]{muller18}
{M{\"u}ller}, O., {Jerjen}, H., \& {Binggeli}, B. 2018, ArXiv e-prints,
  arXiv:1802.08657

\bibitem[{{Munoz} {et~al.}(2018){Munoz}, {Cote}, {Santana}, {Geha}, {Simon},
  {Oyarzun}, {Stetson}, \& {Djorgovski}}]{munoz18}
{Munoz}, R.~R., {Cote}, P., {Santana}, F.~A., {et~al.} 2018, ArXiv e-prints,
  arXiv:1806.06891

\bibitem[{{Nusser}(2018)}]{nusser18}
{Nusser}, A. 2018, ArXiv e-prints, arXiv:1806.01812

\bibitem[{{Peng} {et~al.}(2002){Peng}, {Ho}, {Impey}, \& {Rix}}]{galfit}
{Peng}, C.~Y., {Ho}, L.~C., {Impey}, C.~D., \& {Rix}, H.-W. 2002, \aj, 124, 266

\bibitem[{{Radburn-Smith} {et~al.}(2011){Radburn-Smith}, {de Jong}, {Seth},
  {Bailin}, {Bell}, {Brown}, {Bullock}, {Courteau}, {Dalcanton}, {Ferguson},
  {Goudfrooij}, {Holfeltz}, {Holwerda}, {Purcell}, {Sick}, {Streich}, {Vlajic},
  \& {Zucker}}]{GHOSTS}
{Radburn-Smith}, D.~J., {de Jong}, R.~S., {Seth}, A.~C., {et~al.} 2011, The
  Astrophysical Journal Supplement Series, 195, 18

\bibitem[{{Rizzi} {et~al.}(2007){Rizzi}, {Tully}, {Makarov}, {Makarova},
  {Dolphin}, {Sakai}, \& {Shaya}}]{rizzi}
{Rizzi}, L., {Tully}, R.~B., {Makarov}, D., {et~al.} 2007, \apj, 661, 815

\bibitem[{{Seth} {et~al.}(2005){Seth}, {Dalcanton}, \& {de Jong}}]{seth}
{Seth}, A.~C., {Dalcanton}, J.~J., \& {de Jong}, R.~S. 2005, \aj, 129, 1331

\bibitem[{{Smercina} {et~al.}(2018){Smercina}, {Bell}, {Price}, {D'Souza},
  {Slater}, {Ballin}, {Monachesi}, \& {Nidever}}]{Smercina:2018}
{Smercina}, A., {Bell}, E.~F., {Price}, P.~A., {et~al.} 2018, ArXiv e-prints,
  arXiv:1807.03779

\bibitem[{{Tonry} \& {Schneider}(1988)}]{tonry}
{Tonry}, J., \& {Schneider}, D.~P. 1988, \aj, 96, 807

\bibitem[{{Trentham} \& {Tully}(2002)}]{TT2002}
{Trentham}, N., \& {Tully}, R.~B. 2002, \mnras, 335, 712

\bibitem[{{Trujillo} {et~al.}(2017){Trujillo}, {Roman}, {Filho}, \&
  {S{\'a}nchez Almeida}}]{trujillo2017}
{Trujillo}, I., {Roman}, J., {Filho}, M., \& {S{\'a}nchez Almeida}, J. 2017,
  \apj, 836, 191

\bibitem[{{Tully} {et~al.}(2016){Tully}, {Courtois}, \& {Sorce}}]{Tully:2016}
{Tully}, R.~B., {Courtois}, H.~M., \& {Sorce}, J.~G. 2016, \aj, 152, 50

\bibitem[{{Tully} {et~al.}(2013){Tully}, {Courtois}, {Dolphin}, {Fisher},
  {H{\'e}raudeau}, {Jacobs}, {Karachentsev}, {Makarov}, {Makarova},
  {Mitronova}, {Rizzi}, {Shaya}, {Sorce}, \& {Wu}}]{Tully:2013}
{Tully}, R.~B., {Courtois}, H.~M., {Dolphin}, A.~E., {et~al.} 2013, \aj, 146,
  86

\bibitem[{{van Dokkum} {et~al.}(2018{\natexlab{a}}){van Dokkum}, {Danieli},
  {Cohen}, {Merritt}, {Romanowsky}, {Abraham}, {Brodie}, {Conroy}, {Lokhorst},
  {Mowla}, {O'Sullivan}, \& {Zhang}}]{df2nature}
{van Dokkum}, P., {Danieli}, S., {Cohen}, Y., {et~al.} 2018{\natexlab{a}},
  \nat, 555, 629

\bibitem[{{van Dokkum} {et~al.}(2018{\natexlab{b}}){van Dokkum}, {Cohen},
  {Danieli}, {Romanowsky}, {Abraham}, {Brodie}, {Conroy}, {Kruijssen},
  {Lokhorst}, {Merritt}, {Mowla}, \& {Zhang}}]{df2rn}
{van Dokkum}, P., {Cohen}, Y., {Danieli}, S., {et~al.} 2018{\natexlab{b}},
  Research Notes of the American Astronomical Society, 2, 54

\bibitem[{{van Dokkum} {et~al.}(2018{\natexlab{c}}){van Dokkum}, {Cohen},
  {Danieli}, {Kruijssen}, {Romanowsky}, {Merritt}, {Abraham}, {Brodie},
  {Conroy}, {Lokhorst}, {Mowla}, {O{\textquoteright}Sullivan}, \&
  {Zhang}}]{df2apj}
---. 2018{\natexlab{c}}, \apj, 856, L30

\bibitem[{{van Dokkum}(2001)}]{lacosmic}
{van Dokkum}, P.~G. 2001, \pasp, 113, 1420

\bibitem[{{van Dokkum} {et~al.}(2014){van Dokkum}, {Abraham}, \&
  {Merritt}}]{vDm101}
{van Dokkum}, P.~G., {Abraham}, R., \& {Merritt}, A. 2014, \apj, 782, L24

\bibitem[{{van Dokkum} {et~al.}(2015){van Dokkum}, {Abraham}, {Merritt},
  {Zhang}, {Geha}, \& {Conroy}}]{vd2015a}
{van Dokkum}, P.~G., {Abraham}, R., {Merritt}, A., {et~al.} 2015, \apj, 798,
  L45

\bibitem[{{van Dokkum} {et~al.}(2018{\natexlab{d}}){van Dokkum}, {Danieli},
  {Cohen}, \& {Conroy}}]{vd:2018d}
{van Dokkum}, P.~G., {Danieli}, S., {Cohen}, Y., \& {Conroy}, C.
  2018{\natexlab{d}}, \apjl, LXXX

\bibitem[{{Wasserman} {et~al.}(2018){Wasserman}, {Romanowsky}, {Brodie}, {van
  Dokkum}, {Conroy}, {Abraham}, {Cohen}, \& {Danieli}}]{wasserman18}
{Wasserman}, A., {Romanowsky}, A.~J., {Brodie}, J., {et~al.} 2018, \apjl, LXXX

\bibitem[{{Whiting} {et~al.}(2002){Whiting}, {Hau}, \& {Irwin}}]{WHI}
{Whiting}, A.~B., {Hau}, G. K.~T., \& {Irwin}, M. 2002, The Astrophysical
  Journal Supplement Series, 141, 123

\end{thebibliography}

\end{document}